\documentclass[12pt,aps,prd,showpacs,groupedaddress,notitlepage,footinbib,preprintnumbers]{revtex4-1}
\usepackage{braket,bm}
\usepackage[pass]{geometry} 
\usepackage{color,graphicx}
\usepackage{cancel}
\usepackage{caption}
\usepackage{subfigure}
\usepackage{bm}        
\usepackage{amssymb}   
\usepackage[normalem]{ulem}
\usepackage{mathrsfs,mathtools, wasysym}
\DeclareMathAlphabet{\mathpzc}{OT1}{pzc}{m}{it}

\usepackage{ushort}
\usepackage{accents}
\newcommand{\ul}[1]{\underaccent{\bar}{#1}}
\newcommand{\uc}[1]{\underaccent{\circ}{#1}}
\newcommand{\ub}[1]{\underaccent{\bullet}{#1}}

\newcommand{\D}{\mathrm{d}}

\begin{document}

\preprint{MAN/HEP/2016/01}
\title{Probabilities and signalling in quantum field theory}
\author{Robert Dickinson,}
\email{robert.dickinson-2@manchester.ac.uk}
\author{Jeff Forshaw}
\email{jeff.forshaw@manchester.ac.uk}
\affiliation{Consortium for Fundamental Physics,
  School of Physics and Astronomy,
  University of Manchester,
  Manchester M13 9PL,
  United Kingdom.}
\author{Peter Millington}
\email{p.millington@nottingham.ac.uk}
\affiliation{School of Physics and Astronomy, University of Nottingham, Nottingham NG7 2RD, United Kingdom.}
\date{\today}

\begin{abstract}
We present an approach to computing probabilities in quantum field theory for a wide class of source--detector models. The approach works directly with probabilities and not with squared matrix elements, and the resulting probabilities can be written in terms of expectation values of nested commutators and anti-commutators. We present results that help in the evaluation of these, including an expression for the vacuum expectation values of general nestings of commutators and anti-commutators in scalar field theory. This approach allows one to see clearly how faster-than-light signalling is prevented, because it leads to a diagrammatic expansion in which the retarded propagator plays a prominent role. We illustrate the formalism using the simple case of the much-studied Fermi two-atom problem.  
\end{abstract}


\maketitle

\section{Introduction}

Relativistic quantum field theories respect causality and faster-than-light signalling is forbidden. This well-known fact is a direct consequence of the vanishing of the commutator (or anti-commutator) of field operators when evaluated at spacelike separations (e.g.~see ref.~\cite{Eberhard:1988yj}). It is, however, less clear how faster-than-light signalling (Einstein causality) emerges in explicit calculations, where the Feynman propagator is often ubiquitous.
In this paper, we will develop a means to compute probabilities that resolves this matter in a general way, by highlighting the role of the retarded propagator. The formalism operates at the level of cross-sections and probabilities rather than at the level of amplitudes.

\begin{figure}[h]
\centering
\includegraphics[width=0.8\textwidth]{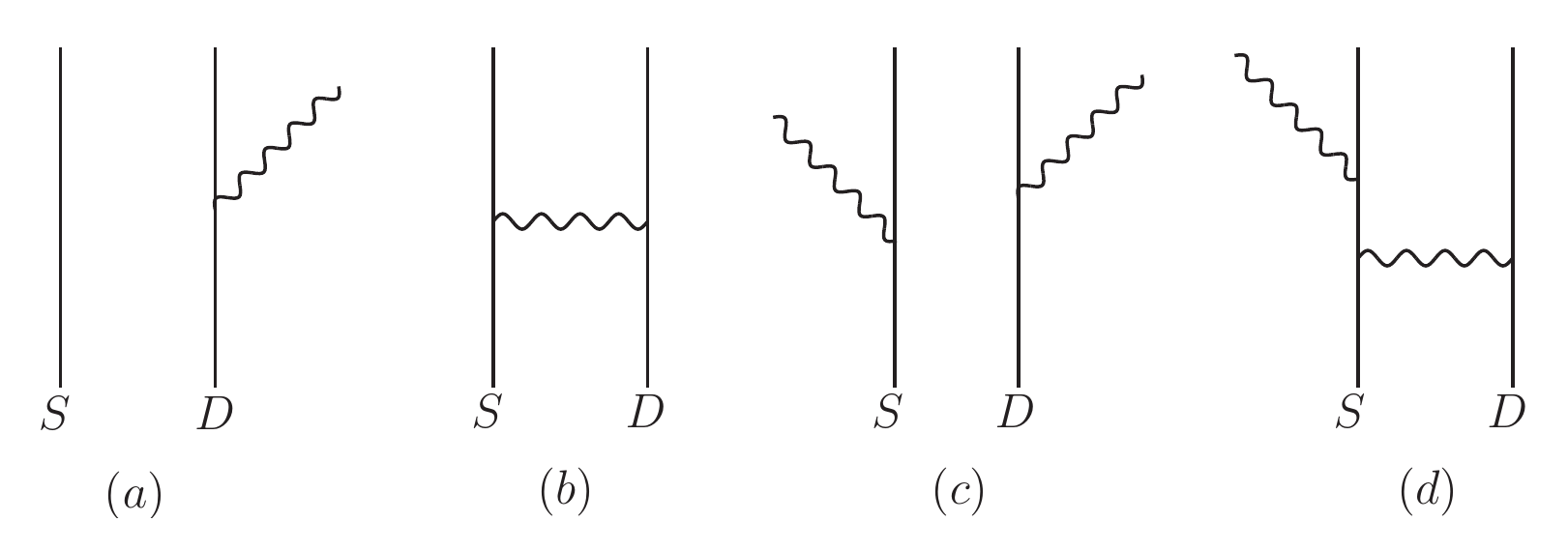}
\caption{The Feynman diagrams corresponding to the amplitudes relevant to the Fermi problem. We show only those graphs that give rise to contributions that depend upon the distance between the source and detector atoms. Solid lines denote the source ($S$) and detector ($D$) atoms and the wavy lines are photons. Time runs upwards.}
\label{fig:f1}
\end{figure}

The archetypal example of a signalling process is the Fermi two-atom problem~\cite{Fermi}. Fermi considered two point-like atoms, A and B, separated by a distance $R$. At time $t=0$, atom A is prepared in an excited state and atom B is prepared in its ground state. He calculated the probability that, at a later time $T$, atom B should be found in its excited state after absorbing a photon emitted during the spontaneous decay of atom A, which ends up in its ground state. Fermi believed this probability should be strictly zero for $T<R/c$, in order to respect Einstein causality, and he claimed to prove it \cite{Fermi}. However, Fermi was wrong \cite{Shirokov1}, for he erroneously approximated an integral over positive frequencies by one over both positive and negative frequencies. The correct result should have been a non-vanishing probability for the excitation of atom B for $T < R/c$.
The history of the Fermi problem is worth recapping and we do so in a footnote \footnote{Fermi's original result was consistent with the previous work of Kikuchi \cite{Kikuchi} and, despite some initial concerns voiced by Ferretti \& Peierls \cite{FP}, was supported by the subsequent work of Heitler \& Ma \cite{Heitler} and Hamilton \cite{Hamilton}. In 1968, two years after Shirokov \cite{Shirokov1}  pointed out Fermi's error, Ferretti essentially solved the Fermi problem \cite{Ferretti}. In particular, he explained the necessity to focus on a local observable with an inclusive sum over unobserved particles. However, Ferretti's work appeared as a chapter in a book and seems not to have been widely appreciated (Shirokov was a noteable exception \cite{Shirokov2}) and, in the 1970s, Fermi's result was still regarded as textbook \cite{Louisell,Milonni}. 
It is ironic that the Ferretti paper starts with the words `In this paper I will not say anything new.' 
In 1987, Rubin \cite{Rubin:1987zz} re-discovered the apparently acausal nature of the two-atom problem but did not explain how it is resolved. In 1990, the correct explanation for the non-violation of Einstein causality in the two-atom problem was re-discovered by Biswas et al. \cite{Biswas} and by Valentini \cite{Valentini}, and by the mid 1990s the dust seems to have settled and the role of the Rotating Wave Approximation in faking causality was appreciated \cite{MilonniJamesFearn,PT,Dolce}. A noteable exception to this was Hegerfeldt's 1994 paper \cite{Hegerfeldt:1993qe}, which created quite a media stir \cite{Maddox,Gribbin} and provoked the clarifying response in \cite{Buchholz:1994eb}.}.    
The fact that atom B is instantaneously correlated with atom A is not a problem for Einstein causality, which is restored if one asks instead for the probability that B is excited at time $T$ with no restriction on the state of atom A or the electromagnetic field, i.e.~if one makes a local measurement on atom B. This is nicely elucidated in the case of heavy atoms and without the complication of renormalization in refs.~\cite{Ferretti,PT}. If one computes the probability that the detector atom is excited at time $T$, regardless of the state of the source atom and the electromagnetic field, then the leading order contributions to the amplitude are illustrated in Figure \ref{fig:f1}. 
Graph (b) is Fermi's and, by itself, it leads to a contribution that does not vanish for $T < R/c$. Adding in the other contributions (graph (c) multiplied by its conjugate and the interference between graphs (a) and (d)) precisely cancels the causality-violating terms. Note that this relies on the fact that an atom in its ground state can fluctuate into an excited state with the emission of a field quantum. This does not violate energy conservation because of Heisenberg's uncertainty principle. Although this treatment involves only bare atomic states, it seems to us that the idea is robust enough to survive renormalization. In this way, superluminal signalling is prevented in the weak sense proved in refs.~\cite{Schlieder,NW,Buchholz:1994eb}. A clear statement of weak causality can be found in ref.~\cite{Hegerfeldt:1998ar}. In essence it says that, although atom B may be excited for any time $T>0$, the excitation probability is independent of the state of atom A if $T < R/c$. For example, suppose Alice, who is located at the source atom, aims to transmit a bit of information to Bob, who is at the detector. To do this, Alice prepares the state of atom A at time zero. Because atom B can be spontaneously excited for any $T>0$, doing this once will not be enough to transmit the bit of information reliably. Alice will need to repeatedly prepare the source atom for each bit she wishes to transmit. Bob will then be able to measure that bit, to a certain statistical precision, by measuring the probability of finding the detector atom to be excited.  

In what follows, we revisit the question of signalling in quantum field theory. Specifically, we will present a new and quite general way to compute probabilities in the interaction picture. This approach makes Einstein causality manifest and has the interesting feature that we do not need to sum explicitly over unobserved emissions. 


\section{A simple source-detector model}
\label{sec:2atom}
In order to develop the formalism in a familiar context, we start by considering two point-like atoms, $S$ and $D$, separated by a distance $R$, which act as source and detector of disturbances in a neutral scalar field, $\phi$. 
In this section, we will present a formalism that allows one to compute the probability of finding the system to be in some particular configuration at time $t=T$ given that it was in some other configuration at time $t=0$. We will consider more general source--detector models in Section \ref{sec:gen}.

We begin by considering a closed system represented by a product of the Hilbert spaces of the source atom, detector atom and field:  $\mathscr{H}=\mathscr{H}^S \times \mathscr{H}^D \times \mathscr{H}^\phi$. For the Hamiltonian, we take $H = H_0 + H_{\text{int}}$, where $H_0 = H_0^S + H_0^D + H_0^\phi$ and $H_\text{int} = H^{S\phi} + H^{D\phi}$. The superscripts refer to the spaces in which the operators act (e.g.~in the case of $H^{S\phi}$, this is the product space $\mathscr{H}^S \times \mathscr{H}^\phi$). In this section, we will only consider interactions between the atoms and the field. Field self-interactions will be considered in Section \ref{sec:gen}.
Under the free part of the Hamiltonian, $H_0$, each atom $X\in\{S,D\}$ has a complete set of bound states $\{\ket{n^X}\}$ with eigenvalues given by $H_0^X\ket{n^X} = \omega^X_n\ket{n^X}$. 

Atoms $S$ and $D$ are assumed to be static and interact with the field at the fixed, spatial points $\mathbf{x}^S$ and $\mathbf{x}^D$ via transition moments $\mu^X_{mn}$, which in this toy scalar field example we will take to be monopole moments. The full interaction-picture Hamiltonian is then
\begin{align}
H_0 &=  \sum_n \omega_n^S \ket{n^S}\bra{n^S} + \sum_n \omega_n^D \ket{n^D}\bra{n^D} + \int \D^3\mathbf{x}\; \Big( \tfrac{1}{2}(\partial_t\phi)^2 + \tfrac{1}{2}(\mathbf{\nabla}\phi)^2 + \tfrac{1}{2}m^2\phi^2 \Big)~, \,\nonumber\\
H_{\mathrm{int}}(t) &= M^S(t)\,\phi(\mathbf{x}^S,t) + M^D(t)\,\phi(\mathbf{x}^D,t) \,, \label{eq:h1ex}
\end{align}
where $M^X(t) \equiv \sum_{mn}\mu_{mn}^X \, e^{i\omega^X_{mn}t} \,\ket{m^X}\bra{n^X}$ and $\omega^X_{mn} = \omega^X_m - \omega^X_n$.
We shall assume that $\mu_{nn}^X=0\;\forall\,n$, i.e.~that emission or absorption of a field quantum always results in a transition up or down in energy. The Fermi problem has also been discussed in the case of two-level (Unruh-DeWitt) point-like detectors in ref.~\cite{Cliche:2009fma} and, for a discussion of potential causality issues in general particle-detector models, see ref.~\cite{Martin-Martinez:2015psa}.

We suppose that the system is initially ($t=0$) described by a density matrix $\rho_0$ and that the measurement outcome is described by an operator $E$. In general, $E$ is an element of a Positive-Operator Valued Measure, and it may be written as a sum over products of hermitian operators:
\begin{align}
E &= \sum_\kappa E^S_{(\kappa)} E^D_{(\kappa)} \mathcal{E}_{(\kappa)} ~.
\end{align}
The superscripts $S$ and $D$ denote the Hilbert space in which the operators act and $\mathcal{E}$ acts in the field Hilbert space.
We explicitly consider a single product, $E = E^S E^D \mathcal{E}$, but the generalization to a sum of such operators is straightforward.
The probability of the measurement outcome, $\mathbb{P}$, is then given by 
\begin{align}
\mathbb{P} &= \mathrm{Tr} (E  \rho_T)~, \label{eq:prob} \\
\rho_T &\equiv U_{T,0}\,\rho_0 \,U^\dag_{T,0}\,  \\
\text{and}\;\;\; U_{T,0} &= \mathrm{T}\exp\Big(\,\tfrac{1}{i}\!\int_0^{T} \! \D t\; H_{\mathrm{int}}(t) \Big)\,.
\end{align}
Note that the measurement is quite general and not restricted to probing only the state of the detector atom. We will consider this restricted case in Section \ref{sec:comexp2}.

One of our goals is to determine the sensitivity of the detector to changes in the preparation of the source. To this end, we will consider an initial mix of two states $\ket{i_p}$ and $\ket{i_g}$:
\begin{equation}
	\rho_0 = \gamma \ket{i_p} \bra{i_p} + (1\!-\!\gamma)\ket{i_g}\bra{i_g}~,
\end{equation}
where
\begin{align}
\ket{i_p} &= \ket{ p^S\,g^D\,0^\phi} \equiv \ket{p^S}\otimes\ket{g^D}\otimes\ket{0^\phi}~~\text{and} ~~
\ket{i_g} = \ket{ g^S\,g^D\,0^\phi} ~.
\label{eq:initialstates}
\end{align}
The first corresponds to the source atom being in an excited state (labelled by $p$) and the detector atom being in its ground state (labelled by $g$), whilst the second corresponds to both the source and detector atoms being in their ground states. In both cases, we suppose that the field is known to have no excitations. Although this is quite a specific initial state, the results that follow can easily be generalized to other initial states. Moreover, in much of what follows the choice of initial state is unimportant. We can define the sensitivity of the detector, $\sigma_{pg}$\,:
\begin{equation}
\sigma_{pg} \equiv \frac{\D \mathbb{P}}{\D\gamma} = \mathbb{P}_p - \mathbb{P}_g  \,,
\label{eq:sensitivity}
\end{equation}
where 
\begin{equation}
	\mathbb{P}_{p,g} \equiv \bra{i_{p,g}} U^\dag_{T,0} E \, U_{T,0} \ket{i_{p,g}}
\end{equation}
is the measurement probability given the state $\ket{i_{p,g}}$ at time $t=0$.
Of course $\mathbb{P}_p$ and $\mathbb{P}_g$  can also be written as squared matrix elements.  However, we do not perform the calculation this way; instead, as in refs.~\cite{Franson,Dickinson:2013lsa,Cliche:2009fma}, we use a generalization of the Baker-Campbell-Hausdorff lemma to commute the operator $E$ through the evolution operator, which gives  
\begin{align}
\mathbb{P}_{p,g} 	&= \sum_{j=0}^\infty \int_0^{T} \D t_1 \D t_2\ldots \D t_j \;\Theta_{12...j} \bra{i_{p,g}}\mathcal{F}_j\ket{i_{p,g}} ~, \label{eq:fsum}
\end{align}
where
\begin{align}
\mathcal{F}_0 &= E \,, \nonumber\\
\mathcal{F}_j &= \tfrac{1}{i}\Big[ \mathcal{F}_{j-1}, H_\mathrm{int}(t_j) \Big] ~,
\label{eq:effs}
\end{align}
and $\Theta_{ijk\ldots} \equiv 1$ if $t_i>t_j>t_k\ldots$ and zero otherwise.
Using the notation $\phi_j^X \equiv \phi(\mathbf{x}^X,t_j)$ and $M^X_j \equiv M^X(t_j)$, we may write
\begin{align}
\mathcal{F}_j &= \tfrac{1}{i}\Big[ \mathcal{F}_{j-1}\,,\, M^S_j\phi^S_j + M^D_j\phi^D_j \Big] ~.
\end{align}
We now show how the $\mathcal{F}_j$ operators can be computed to any order $j$.


\subsection{A general commutator expansion}
\label{sec:comexp}
We start from the following identity for any operators $A^X, B^X \in\mathscr{H}^X$ and $P^\phi,Q^\phi \in\mathscr{H}^\phi$:
\begin{align}
\big[ A^X P^\phi , B^X Q^\phi \big] &\equiv \tfrac{1}{2}\big[ A^X,B^X \big]\big\{ P^\phi, Q^\phi \big\} + \tfrac{1}{2}\big\{ A^X,B^X \big\}\big[ P^\phi, Q^\phi \big]\,,
\end{align}
in which $\{ A^X, B^X \}\equiv A^X B^X + B^X A^X$. With $\mathcal{F}_0=E$, the first commutator is then
\begin{align}
\mathcal{F}_1 &= \tfrac{1}{2i}\big[ E^S , M^S_1 \big] E^D \big\{\mathcal{E},\phi^S_1\big\} + \tfrac{1}{2i}\big\{ E^S , M^S_1 \big\} E^D \big[\mathcal{E},\phi^S_1\big] \nonumber\\
&\;\;\;\;\; + \tfrac{1}{2i}E^S\big[ E^D , M^D_1 \big] \big\{\mathcal{E},\phi^D_1\big\} + \tfrac{1}{2i}E^S\big\{ E^D , M^D_1 \big\} \big[\mathcal{E},\phi^D_1\big] ~.
\end{align}
It will be very convenient to define the following sequences of (hermitian) operators: 
\begin{align}
E^X_{\ldots k} &\equiv \tfrac{1}{i}\big[ E^X_{\ldots}, M^X_k \big]~, & E^X_{\ldots \ul{k}} &\equiv \big\{ E^X_{\ldots}, M^X_k \big\}~,   \nonumber\\
\mathcal{E}^{\ldots X}_{\ldots k} &\equiv \tfrac{1}{i}\big[ \mathcal{E}^{\ldots}_{\ldots}, \phi^X_k \big]~, & \mathcal{E}^{\ldots X}_{\ldots \ul{k}} &\equiv \big\{ \mathcal{E}^{\ldots}_{\ldots}, \phi^X_k \big\}~.
\label{eq:opseqsdef}
\end{align}
Note that the indices on these operators are always time-ordered, with the latest time on the left. Using this notation,
\begin{align}
\mathcal{F}_1 &= \tfrac{1}{2}\big( E^S_1 E^D \mathcal{E}^{S}_{\ul{1}} + E^S_{\ul{1}} E^D \mathcal{E}^{S}_{1} + E^S E^D_1 \mathcal{E}^{D}_{\ul{1}} + E^S E^D_{\ul{1}} \mathcal{E}^{D}_{1} \big) ~.
\label{eq:f1nested}
\end{align}
The $\mathcal{F}_j$ can be expressed in a very compact form
by introducing an under-circle notation,
\begin{equation}
	E_{\uc{k}\uc{l}}\mathcal{E}_{\ub{k}\ub{l}} \equiv E_{kl}\mathcal{E}_{\ul{k}\ul{l}} + E_{k\ul{l}}\mathcal{E}_{\ul{k}l} + E_{\ul{k}l}\mathcal{E}_{k\ul{l}} + E_{\ul{k}\ul{l}}\mathcal{E}_{kl}~,
\end{equation}
which denotes a sum over complementary pairs of commutation operations. Exploiting this notation gives
\begin{align}
\mathcal{F}_0 &= E^S E^D \mathcal{E} ~, \nonumber\\
\mathcal{F}_1 &= \tfrac{1}{2}\big( E^S_{\uc{1}} E^D \mathcal{E}^{S}_{\ub{1}} + E^S E^D_{\uc{1}} \mathcal{E}^{D}_{\ub{1}}  \big) ~, \nonumber\\
\mathcal{F}_2 &= \tfrac{1}{4}\big( 
E^S_{\uc{1}\uc{2}} E^D \mathcal{E}^{SS}_{\ub{1}\ub{2}} + E^S_{\uc{1}} E^D_{\uc{2}} \mathcal{E}^{SD}_{\ub{1}\ub{2}} + E^S_{\uc{2}} E^D_{\uc{1}} \mathcal{E}^{DS}_{\ub{1}\ub{2}} + E^S E^D_{\uc{1}\uc{2}} \mathcal{E}^{DD}_{\ub{1}\ub{2}} 
\big) ~, \nonumber\\
\mathcal{F}_3 &= \tfrac{1}{8}\big(  
E^S_{\uc{1}\uc{2}\uc{3}}E^D\mathcal{E}^{S\!S\!S}_{\ub{1}\ub{2}\ub{3}} 
+ E^S_{(\uc{1}\uc{2}}E^D_{\uc{3})}\mathcal{E}^{(\!S\!S\!D)}_{(\ub{1}\ub{2}\ub{3})} 
+ E^S_{(\uc{1}}E^D_{\uc{2}\uc{3})}\mathcal{E}^{(\!S\!D\!D)}_{(\ub{1}\ub{2}\ub{3})} 
+ E^S E^D_{\uc{1}\uc{2}\uc{3}}\mathcal{E}^{D\!D\!D}_{\ub{1}\ub{2}\ub{3}} 
\big) ~. 
\end{align}
The indices in parentheses in the last line indicate a summation over those permutations of the indices that give rise to unique terms, subject to the  time indices being ordered within each operator. For example, $E^S_{(12}E^D_{3)} = E^S_{12}E^D_3 + E^S_{13}E^D_2 + E^S_{23}E^D_1$. The indices on each $\mathcal{E}^{\ldots}_{\ldots}$ operator are fixed by those on the corresponding product of $E^X_{\cdots}$ operators, i.e.~the $S$ or $D$ label associated with each numerical index matches that of the associated $E^X_{\ldots}$ operator, and its underlining state is complementary to the one it has on $E^X_{\ldots}$. 

The general result for $\mathcal{F}_n$ is extremely simple.
It is the sum of all distinct products of operators of the form $2^{-n} E^S_{\ldots} E^D_{\ldots} \mathcal{E}^{\ldots}_{\ldots}$ with every index $\{1,\ldots,n\}$ appearing once on one of the $E^X_{\ldots}$ and once on $\mathcal{E}^{\ldots}_{\ldots}$:
\begin{align}
\mathcal{F}_n &= 2^{-n}\sum_{a\,=\,0}^{n}E^S_{(\uc{1}\ldots\!\underaccent{\!\!\cdots}{}\; \uc{a}}\,E^D_{a+\!\uc{}\,1\ldots\!\!\!\underaccent{\cdots}{}\;\;\uc{n})}\,\mathcal{E}^{(\!S\ldots S\,\ D\ldots D)}_{(\ub{1}\,\ldots\underaccent{\!\!\!\!\cdots}{}\,\ub{a}\,a+\!\ub{}\,1\ldots\!\!\!\underaccent{\cdots}{}\;\;\ub{n})} ~.
\label{eq:fngen}
\end{align}
In the above summation, the set $i\dots j$ is understood to be the empty set if $i>j$, resulting in a factor of $E^X$ (with no indices).


\subsection{An example: a local measurement}
\label{sec:comexp2}

In this section, we shall focus upon the case of a measurement made only on the $D$ atom, with no restriction on the state of the $S$ atom or the field. Specifically, we compute the probability 
of finding the detector atom in an excited state, $\ket{q^D}$, at time $t=T$.  In this case
\begin{equation}
	E = \sum_{n,\alpha}\ket{ n^S\;q^D\;\alpha^\phi }\bra{ n^S\;q^D\;\alpha^\phi } = \mathbb{I}^S  \ket{q^D}\bra{q^D}~\mathbb{I}^\phi~,  \label{eq:fcase3} 
\end{equation}
where $\mathbb{I}^S$ and $\mathbb{I}^\phi$ are the identity operators in $\mathscr{H}^S$ and $\mathscr{H}^\phi$. 
Notice that we have used the completeness of states to sum over the final states of the source atom and field in Eq.~(\ref{eq:fcase3}). In this way, we avoid ever having to sum explicitly over unobserved final states. This feature of our approach may have interesting consequences for calculations of inclusive observables in S-matrix theory, where, for example, the sum over unobserved emissions is important in securing the cancellation of infra-red singularities in gauge theories. 

Since we fix $\mathcal{E}=\mathbb{I}^\phi$, it follows that
\begin{align}
\mathcal{E}^{X\ldots}_{1\ldots}  &= 0~, & \mathcal{E}^{X}_{\ul{1}} &= 2\phi^X_1 ~, \nonumber\\
\mathcal{E}^{XY}_{\ul{1}{2}} &= \tfrac{2}{i}[ \phi^X_1,\phi^Y_2 ] ~, & \mathcal{E}^{XY}_{\ul{1}\ul{2}} &= 2\{ \phi^X_1,\phi^Y_2 \} ~, \nonumber\\
\mathcal{E}^{XYZ\ldots}_{\ul{1}23\ldots} &= 0 ~.
\label{eq:fieldrelations}
\end{align}
The first of these relations immediately sets half of the terms in Eq.~(\ref{eq:fngen}) to zero and ensures that the `1' index (which labels the latest time) is never underlined on an $E_{1\ldots}^X$ operator. The first and last of the relations in Eq.~(\ref{eq:fieldrelations}) are examples of a more general rule: any $\mathcal{E}^{\ldots}_{\ldots}$ operator vanishes if its first $k$ indices consist of more non-underlined than underlined indices, for any $k$.  
In Appendix~\ref{sec:fieldnests}, we show how to evaluate any $\mathcal{E}^{\ldots}_{\ldots}$ operator and its vacuum expectation value, given $\mathcal{E}=\mathbb{I}^\phi$.

Since we also fix $E^S =\mathbb{I}^S$ (we will consider the case of non-trivial $E^S$ in the next sub-section), it further follows that
\begin{align}
E^S_{k\ldots} &= 0  ~, & E^S_{\ul{k}} &= 2M^S_k ~. \label{eq:snout}
\end{align}
The first of these eliminates half of the remaining terms in Eq.~(\ref{eq:fngen}) and ensures that the first index in $E^S_{\ul{k}\ldots}$ is always underlined. The `1' index must now be carried by the $E^D_{1\ldots}$ operator. 

With these restrictions, up to fourth order, the non-vanishing terms in Eq.~(\ref{eq:fngen}) are
\begin{align}
\mathcal{F}_1 &= \tfrac{1}{2} E^D_1 \mathcal{E}^{D}_{\ul{1}} ~, \nonumber \\
\mathcal{F}_2 &= \tfrac{1}{4}\big(E^D_{12}\mathcal{E}^{D\!D}_{\ul{1}\ul{2}} + E^D_{1\ul{2}}\mathcal{E}^{D\!D}_{\ul{1}2} + E^D_{1}E^S_{\ul{2}}\mathcal{E}^{D\!S}_{\ul{1}2}\big) ~,\nonumber\\
\mathcal{F}_3 &= \tfrac{1}{8}\big(  E^D_{12\uc{3}}\mathcal{E}^{D\!D\!D}_{\ul{1}\ul{2}\ub{3}} +  E^D_{1\ul{2}{3}}\mathcal{E}^{D\!D\!D}_{\ul{1}{2}\ul{3}} + E^D_{12}E^S_{\ul{3}}\mathcal{E}^{D\!D\!S}_{\ul{1}\ul{2}3} + E^D_{13}E^S_{\ul{2}}\mathcal{E}^{D\!S\!D}_{\ul{1}2\ul{3}} + E^D_{1}E^S_{\ul{2}3}\mathcal{E}^{D\!S\!S}_{\ul{1}2\ul{3}} \big) ~, \nonumber\\
\mathcal{F}_4 &= \tfrac{1}{16}\big(
E^D_{1{2}\uc{3}\uc{4}}\mathcal{E}^{D\!D\!D\!D}_{\ul{1}\ul{2}\ub{3}\ub{4}}
+ E^D_{1\ul{2}{3}\uc{4}}\mathcal{E}^{D\!D\!D\!D}_{\ul{1}{2}\ul{3}\ub{4}}
+ E^D_{1{2}\uc{3}}E^S_{\ul{4}}\mathcal{E}^{D\!D\!D\!S}_{\ul{1}\ul{2}\ub{3}4} 
+ E^D_{1\ul{2}{3}}E^S_{\ul{4}}\mathcal{E}^{D\!D\!D\!S}_{\ul{1}{2}\ul{3}4} \nonumber\\&\;\;\;\;\;\;\;\;\;\;
+ E^D_{1{2}\uc{4}}E^S_{\ul{3}}\mathcal{E}^{D\!D\!S\!D}_{\ul{1}\ul{2}3\ub{4}} 
+ E^D_{1{3}\uc{4}}E^S_{\ul{2}}\mathcal{E}^{D\!S\!D\!D}_{\ul{1}2\ul{3}\ub{4}}  
+ E^D_{12}E^S_{\ul{3}\uc{4}}\mathcal{E}^{D\!D\!S\!S}_{\ul{1}\ul{2}3\ub{4}} 
+ E^D_{1{3}}E^S_{\ul{2}\uc{4}}\mathcal{E}^{D\!S\!D\!S}_{\ul{1}2\ul{3}\ub{4}}\nonumber\\&\;\;\;\;\;\;\;\;\;\;
+ E^D_{1\uc{4}}E^S_{\ul{2}{3}}\mathcal{E}^{D\!S\!S\!D}_{\ul{1}2\ul{3}\ub{4}} 
+ E^D_{1}E^S_{\ul{2}{3}\uc{4}}\mathcal{E}^{D\!S\!S\!S}_{\ul{1}2\ul{3}\ub{4}} 
\big) ~.
\end{align}
To compute the measurement probability, we need the expectation values of these operators. (This is where the dependence upon the initial state of the system enters.) In Appendix~\ref{sec:comms}, we present rules to evaluate the expectation values of general atom operators, $E^X_{\cdots}$. For what follows in this section, the following expectation values are useful
\begin{gather}
\bra{g^D}E^D_1\ket{g^D} = 0~, \nonumber\\
\bra{g^D}E^D_{1\ul{2}}\ket{g^D} = |\mu^D_{qg}|^2 \,2\sin \omega^D_{qg}t_{12} ~,~~~
\bra{g^D}E^D_{12}\ket{g^D} = |\mu^D_{qg}|^2 \,2\cos\omega^D_{qg}t_{12}~,  \nonumber \\[0.5em]
\bra{p^S}E^S_{\ul{2}}\ket{p^S} = 2\mu^S_{pp}~, \nonumber\\
\bra{p^S}E^S_{\ul{2}\ul{3}}\ket{p^S}  = \sum_n |\mu^S_{pn}|^2 \,4\cos \omega^S_{pn}t_{23}~, ~~~
\bra{p^S}E^S_{\ul{2}3}\ket{p^S} = \sum_n |\mu^S_{pn}|^2 \,4\sin \omega^S_{pn}t_{23}\,,
\label{eq:evselec}
\end{gather}
where $t_{ij}~\equiv~t_i - t_j$.
For a non-zero contribution to the signal sensitivity $\sigma_{pg}$ with $\mu_{nn}=0$, we require $\bra{g^D}\,E^D_{\ldots}\,\ket{g^D}\neq 0$ and $\bra{p^S}\,E^S_{\ldots}\,\ket{p^S}\neq \bra{g^S}\,E^S_{\ldots}\,\ket{g^S}$, so need only keep terms with at least two indices on each of $E^D_{\ldots}$ and $E^S_{\ldots}$, which means that the first non-zero contribution to $\sigma_{pg}$ arises at fourth order. 

Note that the leading-order contribution to the transition probability $\mathbb{P}_p$ actually comes from $\mathcal{F}_2$:
\begin{align}
\bra{i_p}\mathcal{F}_2\ket{i_p} &= \bra{p^Sg^D 0^\phi} \tfrac{1}{4}\Big(E^D_{12}\mathcal{E}^{D\!D}_{\ul{1}\ul{2}} + E^D_{1\ul{2}}\mathcal{E}^{D\!D}_{\ul{1}2}+ E^D_{1}E^S_{\ul{2}}\mathcal{E}^{D\!S}_{\ul{1}2}\Big)\ket{p^Sg^D 0^\phi}  \nonumber\\
&= |\mu^D_{qg}|^2 \Big(\Delta^{DD(\mathrm{H})}_{12} \!\cos\omega^D_{qg}t_{12} \,+ \, \Delta^{DD(\mathrm{R})}_{12} \!\sin\omega^D_{qg}t_{12} \Big) ~.
\end{align}
However, this does not depend on the state of the source atom and cancels when computing $\sigma_{pg}$. For initial states other than the ones we consider in this section, $\braket{\mathcal{F}_2}$ can contribute to $\sigma_{pg}$. This occurs when the initial density operator $\rho_0$ contains states that are oblique with respect to the projection operator $E^D$, as would be the case if we replaced $\ket{g^D}$ with $\frac{1}{\sqrt{2}}\big(\ket{g^D} + \ket{q^D}\big)$ in $\ket{i_p}$ and $\ket{i_g}$. Additionally, $\braket{\mathcal{F}_3}$ contributes to $\sigma_{pg}$ when $\rho_0$ contains superpositions of field states differing by a single field quantum.

\begin{figure}[h]
\centering
\includegraphics[width=0.87\textwidth]{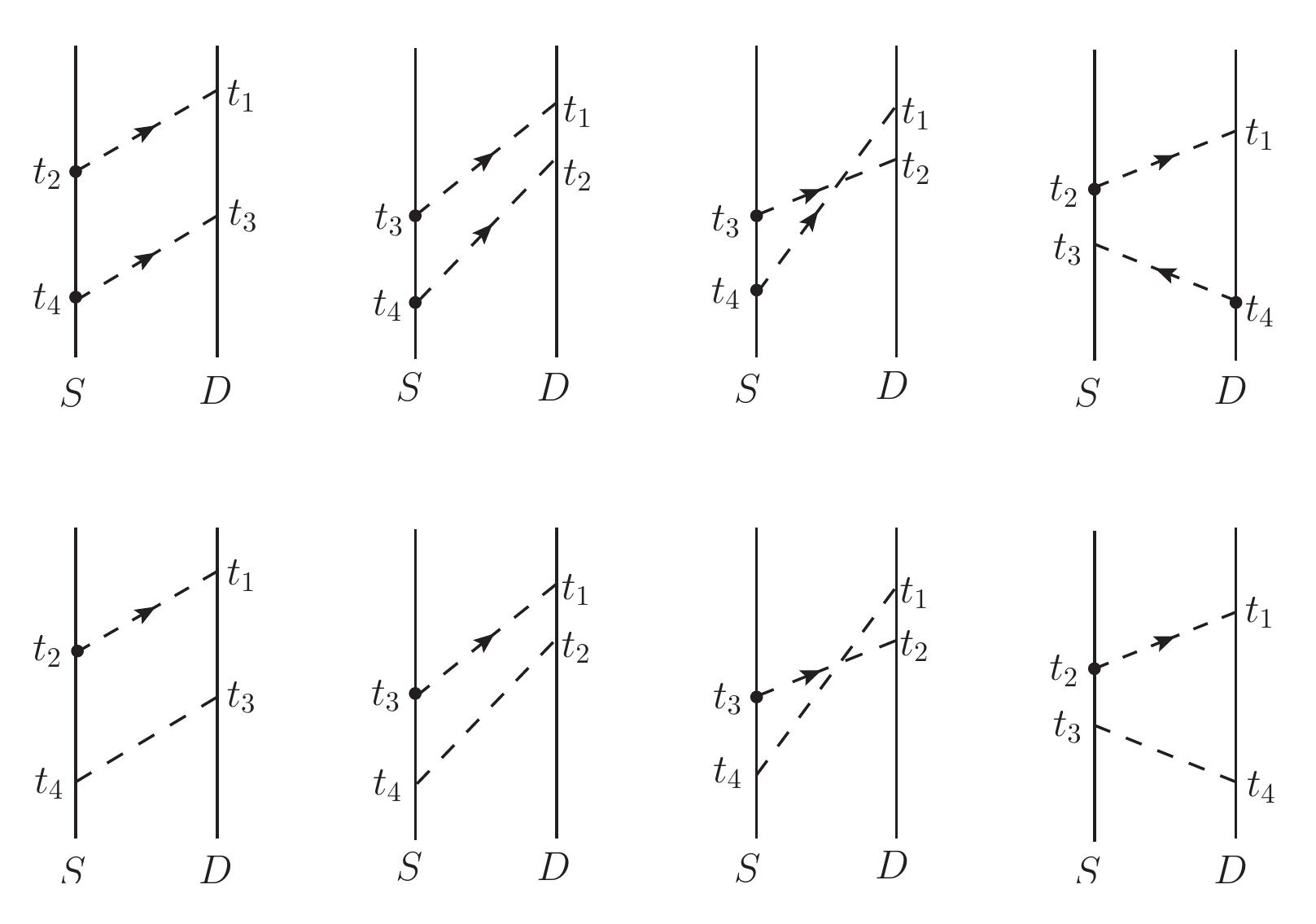}
\caption{\label{fig:f2} The probability-level graphs relevant to the Fermi problem, obtained using the method described in this section. The vertical solid lines denote the source ($S$) and detector ($D$) atoms. The dashed lines with arrows denote retarded propagators, whilst the dashed lines without arrows denote Hadamard propagators. The solid circles on the atom lines indicate the commutator/anti-commutator structure, as discussed in the text.}
\end{figure}

Returning to the calculation of $\sigma_{pg}$, using Eqs.~(\ref{eq:evselec}) and (\ref{eq:fieldvevs}), the contributing terms are
\begin{align}
\bra{i_p}\mathcal{F}_4\ket{i_p} \;&\supset\; \bra{p^S g^D 0^\phi} \;\tfrac{1}{16}\Big( 
E^D_{1{2}}E^S_{\ul{3}\uc{4}}\mathcal{E}^{D\!D\!S\!S}_{\ul{1}\ul{2}3\ub{4}} 
+ E^D_{1{3}}E^S_{\ul{2}\uc{4}}\mathcal{E}^{D\!S\!D\!S}_{\ul{1}2\ul{3}\ub{4}} 
+ E^D_{1\uc{4}}E^S_{\ul{2}{3}}\mathcal{E}^{D\!S\!S\!D}_{\ul{1}2\ul{3}\ub{4}}   
\Big) \;\ket{p^S g^D 0^\phi} \nonumber\\
&= \;\tfrac{1}{16} \braket{E^D_{12}}\Big(\braket{E^S_{\ul{3}{4}}}\braket{\mathcal{E}^{DDSS}_{\ul{1}\ul{2}3\ul{4}}} + \braket{E^S_{\ul{3}\ul{4}}}\braket{\mathcal{E}^{DDSS}_{\ul{1}\ul{2}3{4}}}\Big) \nonumber\\
&\;\;\;\;\;\;\;\;+  \tfrac{1}{16} \braket{E^D_{1{3}}}\Big(\braket{E^S_{\ul{2}{4}}}\braket{\mathcal{E}^{D\!S\!D\!S}_{\ul{1}2\ul{3}\ul{4}}} + \braket{E^S_{\ul{2}\ul{4}}}\braket{\mathcal{E}^{D\!S\!D\!S}_{\ul{1}2\ul{3}{4}}}\Big)  \nonumber\\
&\;\;\;\;\;\;\;\;+  \tfrac{1}{16} \braket{E^D_{1{4}}}\braket{E^S_{\ul{2}{3}}}\braket{\mathcal{E}^{D\!S\!S\!D}_{\ul{1}2\ul{3}\ul{4}}} +  \tfrac{1}{16} \braket{E^D_{1\ul{4}}}\braket{E^S_{\ul{2}{3}}}\braket{\mathcal{E}^{D\!S\!S\!D}_{\ul{1}2\ul{3}{4}}}
\end{align}
\begin{align}
\;\;\;\;&=\; 2 \sum_n |\mu^S_{pn}|^2 \,|\mu^D_{qg}|^2 \Big\{ \cos\omega^D_{qg}t_{12} \Big(  \sin \omega^S_{pn}t_{34} \,\Delta^{DS(\mathrm{H})}_{24}  +  \cos \omega^S_{pn}t_{34} \,\Delta^{DS(\mathrm{R})}_{24} \Big)  \Delta^{DS(\mathrm{R})}_{13} \nonumber\\ 
	 &\;\;\;\;\;\;\;\;\;\;\;\;\;\;\;\;\;\;\;\;\;\;\;\;\;\;\;\;\;\;\;+ \,\cos\omega^D_{qg}t_{12} \Big(  \sin \omega^S_{pn}t_{34} \,\Delta^{DS(\mathrm{H})}_{14}  +  \cos \omega^S_{pn}t_{34} \,\Delta^{DS(\mathrm{R})}_{14} \Big)  \Delta^{DS(\mathrm{R})}_{23}  \nonumber\\[0.5em]
	 &\;\;\;\;\;\;\;\;\;\;\;\;\;\;\;\;\;\;\;\;\;\;\;\;\;\;\;\;\;\;\;+ \cos\omega^D_{qg}t_{13}\, \Big(  \sin \omega^S_{pn}t_{24} \,\Delta^{DS(\mathrm{H})}_{34}  +  \cos \omega^S_{pn}t_{24} \,\Delta^{DS(\mathrm{R})}_{34} \Big)  \Delta^{DS(\mathrm{R})}_{12}  \nonumber\\[0.5em]
	 &\;\;\;\;\;\;\;\;\;\;\;\;\;\;\;\;\;\;\;\;\;\;\;\;\;\;\;\;\;\;\;+ \,\sin \omega^S_{pn}t_{23} \Big(\cos\omega^D_{qg}t_{14} \,\Delta^{SD(\mathrm{H})}_{34}  +\sin \omega^D_{qg}t_{14} \; \Delta^{SD(\mathrm{R})}_{34} \Big) \Delta^{DS(\mathrm{R})}_{12} \Big\} ~. \label{eq:8terms}
\end{align}
The retarded ($\Delta^{XY(\mathrm{R})}_{ij}$) and Hadamard ($\Delta^{XY(\mathrm{H})}_{ij}$) field propagators  are defined in Appendix~\ref{sec:fieldnests}.
We may represent any term in $\braket{\mathcal{F}_n}$ graphically, for arbitary even $n$, using the following rules:
\begin{enumerate}
\item Draw two lines moving forwards in time, corresponding to $S$ and $D$.
\item Draw $n$ vertices associated with the times $t_1$ (latest) to $t_n$ (earliest) 
and distribute them between the two lines with the latest time vertex residing on the $D$ line.
\item If there are vertices on $S$, draw a propagator line between the latest vertex on $S$ and a later vertex on $D$. Pair all other vertices in any combination 
and join each pair with a propagator line.
\item For every vertex at the earlier end of a propagator line, either do nothing or circle the vertex and draw an arrow on the associated propagator. The exception is the latest vertex on the $S$ line, which is always circled and its associated propagator is always arrowed.
\item Associate a factor $\braket{E^X_{ij\ldots}}$ with each line $X$, where $\{i,j,\ldots\}$ is the set of vertices on $X$, and underline every index corresponding to a circled vertex.
\item Associate a factor $\Delta^{XY(\mathrm{R})}_{ij}$ with each arrowed propagator line from $t_i$ on $X$ to $t_j$ on $Y$ 
and a factor $\Delta^{XY(\mathrm{H})}_{ij}$ with each non-arrowed propagator line.
\item Write a factor $(\tfrac{1}{2})^{\frac{n}{2}-1}$, with a further factor of $\tfrac{1}{2}$ if all vertices reside on $D$.
\end{enumerate}
The relations underlying these rules are derived in Appendix \ref{sec:fieldnests}, and  we show the 8 graphs contributing to $\braket{\mathcal{F}_4}$ in Figure \ref{fig:f2}. In general, the number of graphs in $\braket{\mathcal{F}_n}$ is $2^{\frac{n}{2}-1}(n-3)!!(2^{n}+n-3)=\{3,\,34,\,804,\,31320,\ldots\}$ for $n=\{2,4,6,8,\ldots\}$, of which $2^{\frac{n}{2}-1}(n-3)!!\frac{4(n-1)}{n+2} \binom{n-1}{n/2} =\{1,\,8,\,180,\,6720,\ldots\}$ have an equal number of vertices on each line, i.e.~no loops $\Delta^{XX}_{ij}$. 

Crucially, every term in Eq.~(\ref{eq:8terms}) contains a retarded propagator $\Delta^{DS(\mathrm{R})}_{ij}$ with $0 < t_j < t_i < T$, implying that every term in $\sigma_{pg}$ vanishes to fourth order when $T < R$, where $R \equiv |\mathbf{x}^D \!-\! \mathbf{x}^S|$. The stated rules ensure that this holds to all orders. This is in accord with the demands of Einstein causality, i.e.~observation of the detector atom is insensitive to the state of the source atom for times $T < R$. 

In the next section, we will verify that $\sigma_{pg}$ vanishes for spacelike separations to all orders for more general source--detector systems. But, before moving away from the two-atom problem, we will consider the probability of finding, at the time $T$, the detector atom in an excited state ($q^D$) and the source atom in its ground state ($g^S$). Again we will make no restriction on the state of the field at this time. Because this involves measuring the state of two atoms at the same time, it is not local and the probability need not vanish for $T < R$. 

\subsection{An example: a non-local measurement}
\label{sec:simobs}

Now we consider making a measurement on both the source and detector atoms at time $T$. In this case, both $E^D$ and $E^S$ are non-trivial operators. Specifically, we compute the probability of finding atom $D$ in state $\ket{q^D}$ and atom $S$ in state $\ket{g^S}$ without reference to the field, using the following projection operator:
\begin{align}
E &= \sum_{\alpha}\ket{ g^S\;q^D\;\alpha^\phi }\bra{ g^S\;q^D\;\alpha^\phi } = \ket{g^S q^D}\bra{g^S q^D} ~, \label{eq:fcase31}
\end{align}
so that $E^S = \ket{g^S}\bra{g^S}$ and $E^D = \ket{q^D}\bra{q^D}$.

The calculation is similar to the previous case, in that eqs.~(\ref{eq:fieldrelations}) still hold, but eqs.~(\ref{eq:snout}) do not. The non-vanishing terms in Eq.~(\ref{eq:fngen}) are then
\begin{align}
\mathcal{F}_1 &= \tfrac{1}{2}\big( E^S_{1} E^D \mathcal{E}^{S}_{\ul{1}} + E^S E^D_{1} \mathcal{E}^{D}_{\ul{1}}  \big) ~, \nonumber\\
\mathcal{F}_2 &= \tfrac{1}{4}\big( 
E^S_{1\uc{2}} E^D \mathcal{E}^{SS}_{\ul{1}\ub{2}} + E^S_{1} E^D_{\uc{2}} \mathcal{E}^{SD}_{\ul{1}\ub{2}} + E^S_{\uc{2}} E^D_{1} \mathcal{E}^{DS}_{\ul{1}\ub{2}} + E^S E^D_{1\uc{2}} \mathcal{E}^{DD}_{\ul{1}\ub{2}} 
\big) ~,\nonumber\\
\mathcal{F}_3 &= \tfrac{1}{8}\big(  
E^S_{1\ul{2}3}E^D\mathcal{E}^{S\!S\!S}_{\ul{1}2\ul{3}} +E^S_{12\uc{3}}E^D\mathcal{E}^{S\!S\!S}_{\ul{1}\ul{2}\ub{3}}
+ E^S_{({1}\ul{2}}E^D_{3)}\mathcal{E}^{(\!S\!S\!D)}_{(\ul{1}2\ul{3})} + E^S_{({1}{2}}E^D_{\uc{3})}\mathcal{E}^{(\!S\!S\!D)}_{(\ul{1}\ul{2}\ub{3})} \nonumber\\&
+ E^S_{({1}}E^D_{\ul{2}3)}\mathcal{E}^{(\!S\!D\!D)}_{(\ul{1}{2}\ul{3})} + E^S_{({1}}E^D_{{2}\uc{3})}\mathcal{E}^{(\!S\!D\!D)}_{(\ul{1}\ul{2}\ub{3})} 
+ E^S E^D_{{1}\ul{2}{3}}\mathcal{E}^{D\!D\!D}_{\ul{1}{2}\ul{3}} + E^S E^D_{{1}{2}\uc{3}}\mathcal{E}^{D\!D\!D}_{\ul{1}\ul{2}\ub{3}} 
\big)  ~, \nonumber \\
\mathcal{F}_4 &= \tfrac{1}{16}\big(  
E^S_{1\ul{2}3\uc{4}}E^D\mathcal{E}^{S\!S\!S\!S}_{\ul{1}2\ul{3}\ub{4}} + E^S_{12\uc{3}\uc{4}}E^D\mathcal{E}^{S\!S\!S\!S}_{\ul{1}\ul{2}\ub{3}\ub{4}} 
+E^S_{(1\ul{2}3}E^D_{\uc{4})}\mathcal{E}^{(S\!S\!S\!D)}_{(\ul{1}2\ul{3}\ub{4})} +E^S_{(12\uc{3}}E^D_{\uc{4})}\mathcal{E}^{(S\!S\!S\!D)}_{(\ul{1}\ul{2}\ub{3}\ub{4})}\nonumber\\
&+E^S_{(1\ul{2}}E^D_{3\uc{4})}\mathcal{E}^{(S\!S\!D\!D)}_{(\ul{1}2\ul{3}\ub{4})}+E^S_{(12}E^D_{\uc{3}\uc{4})}\mathcal{E}^{(S\!S\!D\!D)}_{(\ul{1}\ul{2}\ub{3}\ub{4})}
+E^S_{(1}E^D_{\ul{2}3\uc{4})}\mathcal{E}^{(S\!D\!D\!D)}_{(\ul{1}2\ul{3}\ub{4})}+E^S_{(1}E^D_{2\uc{3}\uc{4})}\mathcal{E}^{(S\!D\!D\!D)}_{(\ul{1}\ul{2}\ub{3}\ub{4})}\nonumber\\
&+E^SE^D_{1\ul{2}3\uc{4}}\mathcal{E}^{D\!D\!D\!D}_{\ul{1}2\ul{3}\ub{4}}+E^SE^D_{12\uc{3}\uc{4}}\mathcal{E}^{D\!D\!D\!D}_{\ul{1}\ul{2}\ub{3}\ub{4}}
\big) ~.
\end{align}
Notice now that $\mathcal{F}_4$ contains the terms
\begin{align}
\mathcal{F}_4 \;&\supset\; \tfrac{1}{16}\big(  
E^S_{({1}{2}}E^D_{{3}{4})}\mathcal{E}^{(\!S\!S\!D\!D)}_{(\ul{1}\ul{2}\ul{3}\ul{4})} 
+ E^S_{{1}\ul{2}}E^D_{{3}\uc{4}}\mathcal{E}^{\!S\!S\!D\!D}_{\ul{1}{2}\ul{3}\ub{4}} 
+ E^S_{{3}\uc{4}}E^D_{{1}\ul{2}}\mathcal{E}^{\!D\!D\!S\!S}_{\ul{1}{2}\ul{3}\ub{4}} \big) ~.
\end{align}
These terms are noteworthy for the fact that, although they contribute to $\sigma_{pg}$ (since each $E^X_{\cdots}$ operator carries two time indices), they do not give rise to retarded propagators linking $S$ and $D$ (see Appendix \ref{sec:fieldnests}). 
With the initial states specified in Eq.~(\ref{eq:initialstates}), these terms yield the following contributions
to the integrand in Eq.~(\ref{eq:fsum}):
\begin{align}
\bra{i_p}\mathcal{F}_4\ket{i_p} \;&\supset\; \tfrac{1}{4}\braket{E^S_{{1}\ul{2}}}\Delta^{SS(\mathrm{R})}_{12} \Big(\braket{E^D_{{3}{4}}}\Delta^{DD(\mathrm{H})}_{34}
+ \braket{E^D_{{3}\ul{4}}}\Delta^{DD(\mathrm{R})}_{34} \Big)~ \nonumber\\&\;\;\;
+ \tfrac{1}{4}\braket{E^D_{{1}\ul{2}}}\Delta^{DD(\mathrm{R})}_{12} 
\Big(\braket{E^S_{{3}{4}}}\Delta^{SS(\mathrm{H})}_{34} + \braket{E^S_{{3}\ul{4}}}\Delta^{SS(\mathrm{R})}_{34} \Big) \nonumber\\&\;\;\;
+ \tfrac{1}{4}  
\braket{E^S_{({1}{2}}}\braket{E^D_{{3}{4})}}\Delta^{SS(\mathrm{H})}_{(12}\Delta^{DD(\mathrm{H})}_{34)} \nonumber\\&\;\;\;
+ \tfrac{1}{4}  
\braket{E^S_{({1}{2}}}\braket{E^D_{{3}{4})}} \,\Big(\Delta^{SD(\mathrm{H})}_{(13}\Delta^{SD(\mathrm{H})}_{24)} + \Delta^{SD(\mathrm{H})}_{(14}\Delta^{SD(\mathrm{H})}_{23)}\Big) ~.
\end{align}
The first three lines of this expression are independent of the separation $R$
and may be interpreted as representing fluctuations in the state of each atom due to local interactions with the field vacuum. Despite their lack of dependence on $R$, they do depend on the initial source state and so constitute a genuine signal. The terms on the last line give an $R$-dependent contribution to the signal strength,
\begin{align}
\sigma_{pg} \;&\supset\; |\mu^S_{pg}|^2 \,|\mu^D_{qg}|^2 \int^T_0 \D t_i \, \D t_j \, \D t_k \, \D t_l  \;  
 \cos\omega^S_{qp}t_{ij} \, \cos \omega^D_{qg}t_{kl}\;\tfrac{1}{2}\Delta^{SD(\mathrm{H})}_{ik}\Delta^{SD(\mathrm{H})}_{jl}~,
\end{align}
which is non-zero outside the light cone ($T < R$) but falls off approximately exponentially (for massive fields) or as a power (for massless fields) (see Eq.~(\ref{eq:hadtrform})). This may be thought of as a modification to the local interactions between the atoms and the field vacuum that arises from instantaneous vacuum correlations.


\section{More general source--detector models}
\label{sec:gen}
The approach developed in the last section can be used to investigate the influence of one system on another in a significantly more general context. We proceed as before, allowing now for interactions in the fields (which may be bosonic and/or fermionic), i.e.
\begin{equation}
H_\mathrm{int} = H^{D\phi} + H^{S\phi} + H^{\phi\phi} \,.
\end{equation} 
In the interaction picture, a general local interaction of each $X \in \{S,D\}$ with $\phi$ may be written in terms of some series of functions $\widehat{\Phi}^X_\lambda$ of a set of field operators $\phi^a(\mathbf{x},t)$, their spatial derivatives and conjugate momenta $\pi^a(\mathbf{x},t)$: 
\begin{align} \label{eq:general}
H^{X\phi}(t) &= \sum_\lambda \int_{\mathcal{R}^X} \D^3\mathbf{x}\;M^X_\lambda(\mathbf{x},t)\;\widehat{\Phi}^X_\lambda(\phi^a,\nabla\phi^a,\pi^a)~.
\end{align}
Note that each $\widehat{\Phi}^X_\lambda$ is an operator in $\mathscr{H}^\phi$, i.e.~they are not operators in $\mathscr{H}^X$, and each $M^X_\lambda$ is an operator in $\mathscr{H}^X$, which in general varies over the spatial region $\mathcal{R}^X$. We will again use the energy eigenstates of $H_0^X$ as a basis so that
\begin{align}
M^X_\lambda(\mathbf{x},t) \equiv \sum_{m,n}\mu^X_{\lambda,mn}(\mathbf{x}) \,e^{i\omega^X_{mn}t}\ket{m^X}\bra{n^X} \,.
\end{align}
The situation described in Section \ref{sec:2atom} corresponds to that of a single neutral scalar field $\phi$ with a single, localized coupling: $\widehat{\Phi} = \phi$ and $\mu^X_{mn}(\mathbf{x}) = \mu^X_{mn}\delta^{(3)}(\mathbf{x}\!-\!\mathbf{x}^X)$. 

In what follows, we shall restrict ourselves to a single term in the $\lambda$ sum in Eq.~(\ref{eq:general}) and to the case of a single field that interacts via a point-like interaction with $S$ and $D$. The interaction Hamiltonian takes the form
\begin{align}
H_{\mathrm{int}}(t) &= M^S(t)\,\widehat{\Phi}^S(t) + M^D(t)\,\widehat{\Phi}^D(t) + H^{\phi \phi}(t) \, ,
\label{eq:hintgenatom}
\end{align}
where $\widehat{\Phi}^X(t)$ is a local function of the interaction-picture field $\phi^X \equiv \phi(\mathbf{x}^X,t)$ and the conjugate-momentum field $\pi^X \equiv \pi(\mathbf{x}^X,t)$ at the location of system $X$. In the case of fermionic fields, we assume that it is a function of bilinears of the field, i.e.~we do not allow $\widehat{\Phi}$ to carry spinor indices. As discussed in ref.~\cite{Hummer:2015xaa}, this is the case for Unruh-DeWitt detectors. 
This example is similar to the two-atom case in that $S$ and $D$ remain point-like, however we now allow for field self-interactions and for more general interactions between the field and ``atoms.''
Extending to the full Hamiltonian in Eq.~(\ref{eq:general}) should  be straightforward.

Proceeding in the same spirit as Section \ref{sec:2atom}, we write $E = E^S E^D \widehat{\mathcal{E}}$ and introduce the following operators:
\begin{gather}
\widehat{\mathcal{E}}^{\ldots X}_{\ldots k} \equiv \tfrac{1}{i}\big[ \widehat{\mathcal{E}}^{\ldots}_{\ldots}, \widehat{\Phi}^X_k \big]~, ~~~~ \widehat{\mathcal{E}}^{\ldots X}_{\ldots \ul{k}} \equiv \big\{ \widehat{\mathcal{E}}^{\ldots}_{\ldots}, \widehat{\Phi}^X_k \big\}~, ~~~~\widehat{\mathcal{E}}^{\ldots \phi}_{\ldots k} \equiv \tfrac{1}{i}\big[ \widehat{\mathcal{E}}^{\ldots}_{\ldots}, H^{\phi\phi}_k \big]~.
\label{eq:opseqsgen}
\end{gather}
For a general $E$, the first two $\mathcal{F}_n$ operators, defined as in Eq.~(\ref{eq:effs}), are
\begin{align}
\mathcal{F}_1 
&= \tfrac{1}{2i}\big[ E^S , M^S_1 \big] E^D \big\{\widehat{\mathcal{E}},\widehat{\Phi}^S_1\big\} + \tfrac{1}{2i}\big\{ E^S , M^S_1 \big\} E^D \big[\widehat{\mathcal{E}},\widehat{\Phi}^S_1\big] \nonumber\\
&\;\;\;\;\; + \tfrac{1}{2i}E^S\big[ E^D , M^D_1 \big] \big\{\widehat{\mathcal{E}},\widehat{\Phi}^D_1\big\} + \tfrac{1}{2i}E^S\big\{ E^D , M^D_1 \big\} \big[\widehat{\mathcal{E}},\widehat{\Phi}^D_1\big] + \tfrac{1}{i}E^S E^D \big[\widehat{\mathcal{E}},H^{\phi\phi}_1\big] \nonumber\\
&= \tfrac{1}{2}\big( E^S_{\uc{1}} E^D \widehat{\mathcal{E}}^{S}_{\ub{1}} + E^S E^D_{\uc{1}} \widehat{\mathcal{E}}^{D}_{\ub{1}} + 2E^S E^D \widehat{\mathcal{E}}^{\phi}_{1} \big) ~, \nonumber\\[0.3em]
\mathcal{F}_2 &= \tfrac{1}{4}\big( 
E^S_{\uc{1}\uc{2}} E^D \widehat{\mathcal{E}}^{SS}_{\ub{1}\ub{2}} + E^S_{\uc{1}} E^D_{\uc{2}} \widehat{\mathcal{E}}^{SD}_{\ub{1}\ub{2}} + E^S_{\uc{2}} E^D_{\uc{1}} \widehat{\mathcal{E}}^{DS}_{\ub{1}\ub{2}} + E^S E^D_{\uc{1}\uc{2}} \widehat{\mathcal{E}}^{DD}_{\ub{1}\ub{2}} + 2E^S_{\uc{2}} E^D \widehat{\mathcal{E}}^{\phi S}_{1\ub{2}}\nonumber\\
&\;\;\;\;\;\;\;\;\; + 2E^S E^D_{\uc{2}} \widehat{\mathcal{E}}^{\phi D}_{1\ub{2}} + 2E^S_{\uc{1}} E^D \widehat{\mathcal{E}}^{S\phi}_{\ub{1}2} + 2E^S E^D_{\uc{1}} \widehat{\mathcal{E}}^{D\phi}_{\ub{1}2} + 4E^S E^D \widehat{\mathcal{E}}^{\phi\phi}_{12}
\big) ~.
\end{align}
As in Eq.~(\ref{eq:fngen}), we can write down the general result to any order:
\begin{align}
\mathcal{F}_n &= 2^{-n}\sum_{a\,=\,0}^{n}\sum_{b\,=\,a}^{n}2^{b-a}E^S_{(\uc{1}\ldots\!\!\!\underaccent{\:\:\cdots}{}\;\;\uc{a}}\,E^D_{b+\!\uc{}\,1\ldots\!\!\!\underaccent{\cdots}{}\;\;\uc{n})}\,\widehat{\mathcal{E}}^{(\!S\!\ldots S\ \ \phi\ldots\phi\ \; D\ldots D)}_{(\ub{1}\ldots\!\!\!\underaccent{\:\:\cdots}{}\;\;\ub{a}\,a + 1\ldots\!\!\!\;\;b\,b + \!\ub{}\,1\ldots\!\!\!\underaccent{\cdots}{}\;\;\ub{n})} ~.
\label{eq:fngen2}
\end{align}
Compared to Eq.~\eqref{eq:fngen}, we must now sum also over the permutations of terms that contain any number $v\equiv b-a$ of field self-interactions ($0\leq v \leq n$).

To investigate the case of a local measurement, we consider the case of $\widehat{\mathcal{E}}=\mathbb{I}^\phi$ and $\mathcal{E}^{S}=\mathbb{I}^S$. As in Section \ref{sec:comexp2}, we associate the 
non-underlined 1 index with $E^D_{1\ldots}$
and underline the first index on 
$\widehat{\mathcal{E}}^{\ldots}_{\ul{1}\ldots}$ and on $E^S_{\ul{i}\ldots}$:
\begin{align}
\mathcal{F}_1 
&= \tfrac{1}{2} E^D_{1} \widehat{\mathcal{E}}^{D}_{\ul{1}} ~, \nonumber\\
\mathcal{F}_2 &= \tfrac{1}{4}\big( 
E^D_{1\uc{2}} \widehat{\mathcal{E}}^{DD}_{\ul{1}\ub{2}} + E^D_{1} E^S_{\ul{2}} \widehat{\mathcal{E}}^{DS}_{\ul{1}2} + 2 E^D_{1} \widehat{\mathcal{E}}^{D\phi}_{\ul{1}2} 
\big) ~, \nonumber\\
\mathcal{F}_3 &= \tfrac{1}{8}\big(  
E^D_{1\uc{2}\uc{3}}\widehat{\mathcal{E}}^{D\!D\!D}_{\ul{1}\ub{2}\ub{3}} 
+ E^D_{1\uc{2}}E^S_{\ul{3}}\widehat{\mathcal{E}}^{D\!D\!S}_{\ul{1}\ub{2}3} 
+ E^D_{1\uc{3}}E^S_{\ul{2}}\widehat{\mathcal{E}}^{D\!S\!D}_{\ul{1}2\ub{3}} 
+ E^D_{1}E^S_{\ul{2}\uc{3}}\widehat{\mathcal{E}}^{D\!S\!S}_{\ul{1}2\ub{3}}
+ 2E^D_{1\uc{2}} \widehat{\mathcal{E}}^{DD\phi}_{\ul{1}\ub{2}3}  \nonumber\\&\;\;\;\;\;\;\;\;\;\;
+ 2 E^D_{1\uc{3}} \widehat{\mathcal{E}}^{D\phi D}_{\ul{1}{2}\ub{3}}
+ 2E^D_{1} E^S_{\ul{2}} \widehat{\mathcal{E}}^{DS\phi}_{\ul{1}2{3}}
+ 2 E^D_{1}E^S_{\ul{3}} \widehat{\mathcal{E}}^{D\phi S}_{\ul{1}{2}{3}}
+ 4 E^D_{1} \widehat{\mathcal{E}}^{D\phi \phi}_{\ul{1}{2}{3}} 
\big) ~.
\end{align}
Note that when $\widehat{\Phi}$ is linear in $\phi$ or its derivatives, we can
eliminate $\widehat{\mathcal{E}}^{\dots}_{\dots}$ operators whose first $k$ indices before a field self-interaction contain more non-underlined
than underlined indices for any $k$.
For initial states that are indistinguishable by any measurement on $D$ and $\phi$, the only terms making a non-zero contribution to the sensitivity of the detector are those containing an $E^S_{\ul{k}\ldots}$ operator and therefore at least one non-underlined $S$ index on $\widehat{\mathcal{E}}^{\ldots \phi \ldots}_{\ldots k\ldots}$. 

Every non-underlined index $k$ will be involved in a commutator with some index $j$ to its left, representing a later time. In contrast to the simpler case, where index $j$ would always be underlined, $j$ may now be any kind of index. If it is underlined, it generates one of the following four commutators: $[ \,\widehat{\Phi}_j\,,\,\widehat{\Phi}_k \,]$, $[ \,H^{\phi\phi}_j\,,\,\widehat{\Phi}_k \,]$, $[ \,\widehat{\Phi}_j \,,\,H^{\phi\phi}_k\,]$ or $[ \,H^{\phi\phi}_j\,,\,H^{\phi\phi}_k \,]$. Each of these necessarily has the field commutator $[ \phi_j,\phi_k ]$ (or one containing its derivatives) as a factor, ensuring the presence of a retarded propagator $\Delta^{(\mathrm{R})}_{jk}$.
If index $j$ is not underlined, it will instead be involved in a nested commutator such as $\big[\,[ \,\ldots\,,\,\widehat{\Phi}_j \,]\,,\widehat{\Phi}_k\big]$ along with one or more indices to its left, until an underlined index is reached, e.g.~if the next index $i$ in this chain is underlined, we obtain the double nested commutator $\big[\,[ \,\widehat{\Phi}_i\,,\,\widehat{\Phi}_j \,]\,,\widehat{\Phi}_k\big]$. This object may be written as a sum of two terms, one of which has $[ \phi_i,\phi_j ][ \phi_i,\phi_k ]$ as a factor and the other has the two-commutator chain $[ \phi_i,\phi_j ][ \phi_j,\phi_k ]$, ensuring the presence of either $\Delta^{(\mathrm{R})}_{ik}$ or $\Delta^{(\mathrm{R})}_{ij}\Delta^{(\mathrm{R})}_{jk}$.
This pattern of generating retarded propagators continues to any level of nesting, and guarantees the presence, as a factor, of a retarded propagator from every non-underlined index in a chain to at least one vertex later in that chain \cite{Dickinson:2013lsa}. Any such chain must terminate on an underlined index of $\widehat{\mathcal{E}}^{\dots D\dots}_{\dots\ul{k}\dots}$, which is necessarily complemented by an operator $E_{\dots k \dots}^D$. This includes the chain beginning on the latest vertex of $S$. Therefore, we know that the latest  vertex on $S$ must always be connected to some later vertex on $D$  by an unbroken chain of retarded propagators, and hence that all such terms vanish to all orders when $T<R$. In other words, any detector sensitivity will vanish to all orders if the source and detector are spacelike separated. This result continues to hold for any number of interacting fields and to extended sources and detectors.

We close by remarking that the probabilities we compute resemble in-in expectation values. Encouraged, in addition, by the rules articulated in Appendix \ref{sec:comms}, we anticipate that they could be derived from a path-integral approach based upon the Schwinger-Keldysh closed-time path (CTP) formalism~\cite{Schwinger:1960qe,Keldysh:1964ud} and in the spirit of ref.~\cite{Dickinson:2013lsa}. We suspect that we may well be able to generate these probabilities by means of the Kobes-Semenoff cutting rules~\cite{Kobes:1985kc,Kobes:1986za} (the Cutkosky rules~\cite{Cutkosky:1960sp,'tHooft:1973pz} of the CTP formalism), which are known to deliver retarded functions~\cite{Kobes:1990ua} (see also ref.~\cite{Dickinson:2013lsa}).


\appendix

\newpage

\section{Nested commutators and anti-commutators of field operators}
\label{sec:fieldnests}

In this appendix, we provide rules to evaluate general nested commutators and anti-commutators for a scalar field $\phi$, and their vacuum expectation values. We introduce $i\Delta_{ij} \equiv [\phi_i,\phi_j]$ and use $\phi_{(1}\phi_2\ldots\phi_{n)}$ to denote the completely symmetric sum of products of $n$ field operators. 
We also make use of the Feynman, retarded and Hadamard propagators defined as follows: 
\begin{align}
\Delta^{XY(\mathrm{F})}_{ij} &\equiv \bra{0}\,\mathrm{T}\big(\phi^X_i\phi^Y_j\big)\,\ket{0}~, \nonumber\\
\Delta^{XY(\mathrm{R})}_{ij} &\equiv \Theta_{ij}\bra{0}\,\tfrac{1}{i}\big[\phi^X_i,\phi^Y_j\big]\,\ket{0} = \Theta_{ij} \; 2\,\text{Im} \big(\Delta^{XY(\mathrm{F})}_{ij}\big)~, \nonumber\\
\Delta^{XY(\mathrm{H})}_{ij} &\equiv \bra{0}\,\big\{\phi^X_i,\phi^Y_j\big\}\,\ket{0} \;\;\;\;= 2\, \text{Re} \big(\Delta^{XY(\mathrm{F})}_{ij}\big) \,.
\label{eq:propdefs}
\end{align}
Factors of $i$ have been chosen such that $\Delta^{XY(\mathrm{R})}_{ij}$ and $\Delta^{XY(\mathrm{H})}_{ij}$ are real-valued distributions, e.g. see Eq.~(\ref{eq:hadtrform}).

We use an induction argument to prove the following result for the nested anti-commutators defined in Eq.~(\ref{eq:opseqsdef}):
\begin{align}
\mathcal{E}_{\ul{1}\ul{2}\ldots\ul{n}} = \frac{2^n}{n!}\phi_{(1}\phi_2\ldots\phi_{n)} ~.
\label{eq:symresult}
\end{align}
Taking the anti-commutator of both sides of Eq.~(\ref{eq:symresult}) with $\phi_{n+1}$, we obtain
\begin{align}
\mathcal{E}_{\ul{1}\ul{2}\ldots\ul{n}\,\ul{n\!+\!1}} &= \frac{2^n}{n!}\Big(\phi_{(1}\phi_2\ldots\phi_{n)}\phi_{n+1} \;+\; \phi_{n+1}\phi_{(1}\phi_2\ldots\phi_{n)}\Big) ~.
\label{eq:syminduct}
\end{align}
Now we consider $n+1$ copies of the expression on the right and consider the first term in the $r${-}th copy. In this term, commute $\phi_{n+1}$ through $r-1$ places to the left, picking up $r-1$ commutator pieces of the form $i\Delta_{k\,n\!+\!1}\;\phi_{(1}\ldots\phi_{k-1}\phi_{k+1}\ldots\phi_{n)}$. For the second term in the same copy, we commute $\phi_{n+1}$ through $r-1$ places to the right, picking up identical commutator pieces but of opposite sign. Summing these copies of Eq.~(\ref{eq:syminduct}), we have
\begin{align}
(n+1)\mathcal{E}_{\ul{1}\ul{2}\ldots\ul{n}\,\ul{n\!+\!1}} &= \frac{2^n}{n!}\Big(\phi_{(1}\phi_2\ldots\phi_n\phi_{n+1)} \;+\; \phi_{(1}\phi_2\ldots\phi_n\phi_{n+1)}\Big)~.
\end{align}
Since $\mathcal{E} = \mathbb{I}^\phi$, result (\ref{eq:symresult}) is proven for all $n\geq 0$.

We will now derive a result based on Wick's theorem, which is useful when calculating the expectation value of the nested anti-commutators. We begin with the case $n=2$:\begin{align}
\phi_1 \phi_2 &= \;: \phi_1\phi_2 :\, + ~\Delta^{(\mathrm{F})}_{12}~, & 
\phi_2 \phi_1 &= \;: \phi_1\phi_2 :\,  + ~\Delta^{(\mathrm{F})*}_{12} ~.
\label{eq:wicks}
\end{align}
Colons indicate normal ordering of operators, and it is understood that $t_1>t_2>\ldots>t_n$. If we introduce more fields into the product, with any given time-ordering, we generate contractions between pairs of fields in accordance with Eq.~(\ref{eq:wicks}). For example, 
\begin{align}
\phi_1 \phi_3\phi_2 &= \;: \phi_1\phi_2\phi_3:\, +\;\Delta^{(\mathrm{F})}_{12}\phi_3 +\,\Delta^{(\mathrm{F})}_{13}\phi_2 + \;\Delta^{(\mathrm{F})*}_{23}\phi_1 ~, \nonumber\\
\phi_1 \phi_4\phi_3\phi_2 &= \;: \phi_1\phi_2\phi_3\phi_4:\, +\;\Delta^{(\mathrm{F})}_{12}:\phi_3\phi_4: +\,\Delta^{(\mathrm{F})}_{13}:\phi_2\phi_4:+\;\Delta^{(\mathrm{F})}_{14}:\phi_2\phi_3: + \;\Delta^{(\mathrm{F})*}_{23}:\phi_1\phi_4: \nonumber\\
&\;\;\;\;\;+\;\Delta^{(\mathrm{F})*}_{24}:\phi_1\phi_3:  +\;\Delta^{(\mathrm{F})*}_{34}:\phi_1\phi_2: + \;\Delta^{(\mathrm{F})}_{12}\Delta^{(\mathrm{F})*}_{34} + \Delta^{(\mathrm{F})}_{13}\Delta^{(\mathrm{F})*}_{24} + \Delta^{(\mathrm{F})}_{14}\Delta^{(\mathrm{F})*}_{23}~.
\end{align}
In general, this is a sum over all distinct permutations of indices amongst the propagators and normal-ordered products of fields. If we now sum over all orderings of $n$ fields on the left-hand side, only the real parts of each propagator remain:
\begin{align}
\phi_{(1}\phi_2\ldots\phi_{n)} =  n!:\phi_1\phi_2\ldots \phi_n: &+\;n!\,\mathrm{Re}\big(\Delta^{(\mathrm{F})}_{12}\big):\phi_3\ldots\phi_n: + \ldots \nonumber\\
&+ \;n! \,\mathrm{Re}\big(\Delta^{(\mathrm{F})}_{12}\big)\mathrm{Re}\big(\Delta^{(\mathrm{F})}_{34}\big):\phi_5\ldots\phi_n: + \ldots
\end{align}
Using Eq.~(\ref{eq:propdefs}), this yields a general expression for the vacuum expectation of a completely symmetric sum of products of $n$ field operators:
\begin{align}
\bra{0} \phi_{(1}\phi_2\ldots\phi_{n)} \ket{0} &= \begin{cases}
\dfrac{n!}{2^{n/2}}\sum \Delta^{(\mathrm{H})}_{a_1 a_2}\ldots \Delta^{(\mathrm{H})}_{a_{n\!-\!1} a_n} &\text{ if $n$ is even} \\
0 &\text{ if $n$ is odd} \end{cases}
\label{eq:genexpval}
\end{align}
where the sum is over all distinct pairings of indices from the set $\{1,\ldots,n\}$.

Finally, we state an interesting result for a general nesting of commutators and anti-commutators of field operators, with examples given below. Using Eq.~\eqref{eq:symresult}, any operator $\mathcal{E}_{\ldots}$, as defined in Eq.~(\ref{eq:opseqsdef}), with $n$ indices associated with times $t_1>t_2>\ldots > t_n$ and any combination of underlinings, can be written
\begin{align}
\mathcal{E}_{\ldots} &= \sum \,2^r \,\Delta_{a_1 b_1}\ldots\Delta_{a_r b_r}\, \frac{2^s}{s!}\phi_{(c_1}\ldots\phi_{c_s)} ~,
\label{eq:curlye}
\end{align}
where $\{b_i\}$ is the set of $r$ non-underlined indices on $\mathcal{E}_{\ldots}$, each paired with an underlined index $a_i<b_i$ and $\{c_j\}$ is the set of $s=n\!-\!2r$ unpaired indices. Applying Eq.~\eqref{eq:genexpval}, this has the following implication:
\begin{quote}
	The vacuum expectation value of a general nesting of commutators and anti-commutators, i.e. $\mathcal{E}_{1\ldots (2p)}$ with any combination of underlinings, can be written as $2^p$ times the sum of all distinct products of $p$ propagators subject to the following rule: every non-underlined (commutation) index must become the second index on a retarded propagator and all remaining indices are paired and associated with Hadamard propagators.
\end{quote}
In particular, $\bra{0}\mathcal{E}_{\ul{1}2}\ket{0} = 2\Delta^{(\mathrm{R})}_{12}$, $\bra{0}\mathcal{E}_{\ul{1}\ul{2}}\ket{0} = 2\Delta^{(\mathrm{H})}_{12}$,
\begin{eqnarray}
\bra{0}\mathcal{E}_{\ul{1}2\ul{3}4}\ket{0} &=& \bra{0}4\Delta_{12}\Delta_{34} \ket{0} = 4\Delta^{(\mathrm{R})}_{12}\Delta^{(\mathrm{R})}_{34} ~,\nonumber\\
\bra{0}\mathcal{E}_{\ul{1}2\ul{3}\ul{4}}\ket{0}  &=& \bra{0}4\Delta_{12}\phi_{(3}\phi_{4)}\ket{0} = 4\Delta^{(\mathrm{R})}_{12}\Delta^{(\mathrm{H})}_{34} ~,\nonumber\\
\bra{0}\mathcal{E}_{\ul{1}\ul{2}34}\ket{0} &=& \bra{0} 4\big(\Delta_{13}\Delta_{24} + \Delta_{23}\Delta_{14}\big)\ket{0} =  4\big(\Delta^{(\mathrm{R})}_{13}\Delta^{(\mathrm{R})}_{24} + \Delta^{(\mathrm{R})}_{23}\Delta^{(\mathrm{R})}_{14}\big) ~,\nonumber\\
\bra{0}\mathcal{E}_{\ul{1}\ul{2}3\ul{4}}\ket{0}  &=& \bra{0} 4\big(\Delta_{13}\phi_{(2}\phi_{4)} + \Delta_{23}\phi_{(1}\phi_{4)}\big)\ket{0} =  4\big(\Delta^{(\mathrm{R})}_{13}\Delta^{(\mathrm{H})}_{24} + \Delta^{(\mathrm{R})}_{23}\Delta^{(\mathrm{H})}_{14}\big) ~,\nonumber\\
\bra{0}\mathcal{E}_{\ul{1}\ul{2}\ul{3}4}\ket{0} &=& \bra{0} 4\big(\phi_{(1}\phi_{2} \Delta_{3)4} \big) \ket{0} =  4\big(\Delta^{(\mathrm{H})}_{12}\Delta^{(\mathrm{R})}_{34} + \Delta^{(\mathrm{H})}_{13}\Delta^{(\mathrm{R})}_{24} + \Delta^{(\mathrm{H})}_{23}\Delta^{(\mathrm{R})}_{14}\big) ~,\nonumber\\
\bra{0}\mathcal{E}_{\ul{1}\ul{2}\ul{3}\ul{4}}\ket{0} &=& \bra{0} \tfrac{2}{3}\phi_{(1}\phi_{2}\phi_{3}\phi_{4)} \ket{0} =  4\big(\Delta^{(\mathrm{H})}_{12}\Delta^{(\mathrm{H})}_{34} + \Delta^{(\mathrm{H})}_{13}\Delta^{(\mathrm{H})}_{24} + \Delta^{(\mathrm{H})}_{23}\Delta^{(\mathrm{H})}_{14}\big) ~.
\label{eq:fieldvevs}
\end{eqnarray}
In the specific case of two atoms at fixed locations, with $R \equiv |\mathbf{x}^D-\mathbf{x}^S|$ and $z_{ij}~\equiv~m\sqrt{|t_{ij}^2-R^2|}$, the propagators are given by
\begin{align}
\Delta^{DS(\mathrm{R})}_{ij} &= -\frac{\delta(t_{ij}\!-\!R)}{4\pi R} + \frac{m^2}{4\pi}\frac{J_1(z_{ij})}{z_{ij}}\Theta(t_{ij}\!-\!R) ~,\nonumber\\
\Delta^{DS(\mathrm{H})}_{ij} &= -\frac{m^2}{4\pi}\frac{Y_1(z_{ij})}{z_{ij}}\Theta(t_{ij}^2\!-\!R^2) - \frac{m^2}{2\pi^2}\frac{K_1(z_{ij})}{z_{ij}}\Theta(R^2\!-\!t_{ij}^2) = \frac{1}{2\pi(t_{ij}^2\!-\!R^2)} +\mathcal{O}(m^2) ~, \nonumber\\
&\rightarrow -m^2e^{-z_{ij}}(2\pi z_{ij})^{-3/2} ~\text{in the spacelike limit when }m \neq 0\,.
\label{eq:hadtrform}
\end{align}
$J_1, Y_1,$ and $K_1$ are the usual Bessel functions.


\section{Nested commutators and anti-commutators of atom operators}
\label{sec:comms}

In this appendix, we provide rules for obtaining the expectation values $\braket{E_{ij\dots}^X}$. We use the following notation for commutators and anti-commutators:
\begin{equation}
[A,B]_\eta \equiv AB + \eta BA\;,
\end{equation}
with $\eta\in\{+1,-1\}$. In addition, we leave superscripts $X$ and sums over state indices $m,n,r,s,\dots$ implicit. Note that what follows cannot be obtained using the results of Appendix~\ref{sec:fieldnests} for two reasons: firstly, the commutator of two atom operators is not proportional to the unit operator, as is the case for the scalar field; and, secondly, a measurement is involved such that $E^D$ and $E^S$ may not be the identity.

The sequence of commutators or anti-commutators developed from a general hermitian operator $E \equiv \epsilon_{mn}\ket{m}\!\bra{n}$, where the $\epsilon_{mn}$ are constants, is as follows:
\begin{align}
E &= \epsilon_{mn}\ket{m}\!\bra{n} ~,\nonumber\\
\big[E,M_i\big]_{\eta_i} &= \Big(\epsilon_{ms}\mu_{rn}e^{i\omega_{r}t_i}e^{-i\omega_n t_i}+\eta_i\big(m\leftrightarrow r,n\leftrightarrow s\big)\Big)\delta_{sr} \ket{m}\!\bra{n}~, \nonumber\\
\big[\big[E,M_i\big]_{\eta_i} ,M_j\big]_{\eta_j} &= \Big[\Big(\epsilon_{ms}\mu_{ru}\mu_{tn}e^{i\omega_{r}t_i}e^{-i\omega_u t_i}e^{i\omega_tt_j}e^{-i\omega_nt_j}+\eta_i\big(m\leftrightarrow r,s\leftrightarrow u\big)\Big)~\nonumber\\&\qquad+\eta_j\big(m\leftrightarrow t,u\leftrightarrow n\big)\Big]\delta_{sr}\,\delta_{ut} \ket{m}\!\bra{n} ~,\nonumber\\
\big[\big[\big[E,M_i\big]_{\eta_i} ,M_j\big]_{\eta_j},M_k\big]_{\eta_k} & = \Big\{\Big[\Big(\epsilon_{ms}\mu_{ru}\mu_{tw}\mu_{vn}e^{i\omega_rt_i}e^{-i\omega_ut_i}e^{i\omega_tt_j}e^{-i\omega_wt_j}e^{i\omega_vt_k}e^{-i\omega_nt_k}\nonumber\\&\qquad +\eta_i\big(m\leftrightarrow r, s\leftrightarrow u\big)\Big)+\eta_j\big(m\leftrightarrow t,u\leftrightarrow w\big)\Big]\nonumber\\&\qquad +\eta_k\big(m\leftrightarrow v,w\leftrightarrow n\big)\Big\}\delta_{sr}\,\delta_{ut}\,\delta_{wv}\ket{m}\!\bra{n}~.
\end{align}
In the case $E = \ket{q}\!\bra{q}$, the only non-zero terms in the ground state expectation value of the sequence above have $m=n=g$: 
\begin{align}
\label{eq:atomcomm}
\bra{g}\big[E,M_i\big]_{\eta_i}\ket{g} &= 0 ~,\nonumber\\
\bra{g}\big[\big[E,M_i\big]_{\eta_{i}} ,M_j\big]_{\eta_j}\ket{g} &= \eta_i\mu_{gq}\mu_{qg}\Delta^{q(>)}_{ij}\Delta^{g(>)}_{ji}+\eta_j\mu_{qg}\mu_{gq}\Delta^{g(>)}_{ij}\Delta^{q(>)}_{ji}~,\nonumber\\
\bra{g}\big[\big[\big[E,M_i\big]_{\eta_i} , M_j\big]_{\eta_j},M_k\big]_{\eta_k} \ket{g} 
&= \eta_i \mu_{gq}\mu_{qr}\mu_{rg}\Delta^{q(>)}_{ij}\Delta^{r(>)}_{jk}\Delta^{g(>)}_{ki}\nonumber\\&\qquad
+ \eta_j \mu_{gq}\mu_{qr}\mu_{rg}\Delta^{q(>)}_{ji}\Delta^{r(>)}_{ik}\Delta^{g(>)}_{kj}\nonumber\\&\qquad
+\eta_k \mu_{gq}\mu_{qr}\mu_{rg}\Delta^{q(>)}_{ki}\Delta^{r(>)}_{ij}\Delta^{g(>)}_{jk}\nonumber\\&\qquad
+\eta_i\eta_j \mu_{gr}\mu_{rq}\mu_{qg}\Delta^{g(>)}_{kj}\Delta^{r(>)}_{ji}\Delta^{q(>)}_{ik}\nonumber\\&\qquad
+\eta_i\eta_k \mu_{gr}\mu_{rq}\mu_{qg}\Delta^{g(>)}_{jk}\Delta^{r(>)}_{ki}\Delta^{q(>)}_{ij}\nonumber\\&\qquad
+\eta_j\eta_k \mu_{gr}\mu_{rq}\mu_{qg}\Delta^{g(>)}_{ik}\Delta^{r(>)}_{kj}\Delta^{q(>)}_{ji}~,
\end{align}
where $\Delta^{r(>)}_{ij} = e^{-i \omega_r t_{ij}}$ is the (positive-frequency) atom Wightman propagator. Note that $\Delta^{r(>)}_{ij}=\Delta^{r(\mathrm{F})}_{ij}$ (the atom Feynman propagator) when $t_i>t_j$, and $\Delta^{r(>)}_{ij}=\Delta^{r(\mathrm{F})*}_{ij}$ (the atom Dyson propagator) when $t_i<t_j$.

\begin{figure}
\subfigure[]{\includegraphics[]{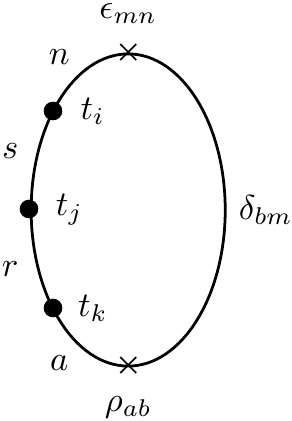}}\hspace{1em}
\subfigure[]{\includegraphics[]{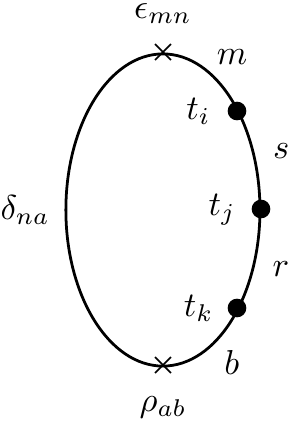}}\hspace{1em}
\subfigure[]{\includegraphics[]{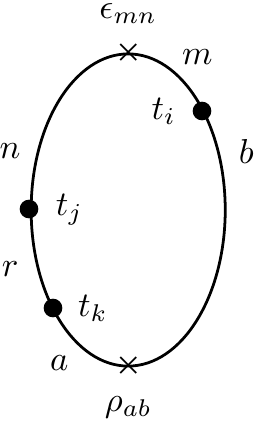}}\hspace{1em}
\subfigure[]{\includegraphics[]{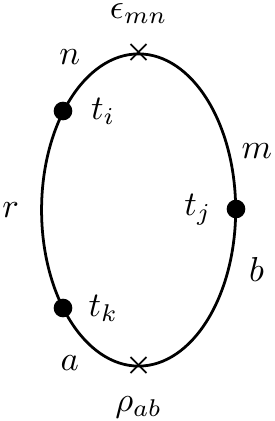}}\\
\subfigure[]{\includegraphics[]{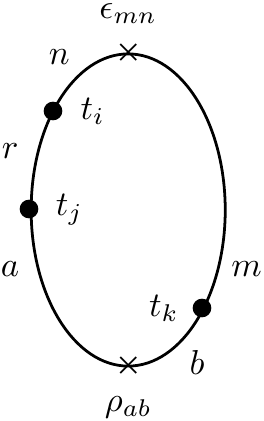}}\hspace{1em}
\subfigure[]{\includegraphics[]{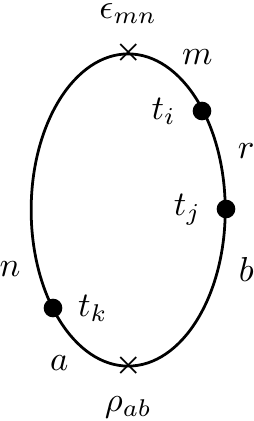}}\hspace{1em}
\subfigure[]{\includegraphics[]{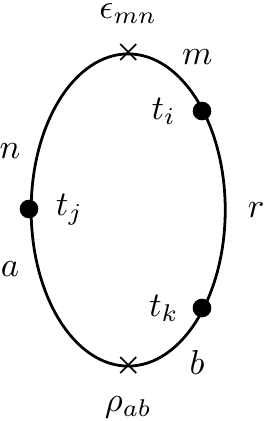}}\hspace{1em}
\subfigure[]{\includegraphics[]{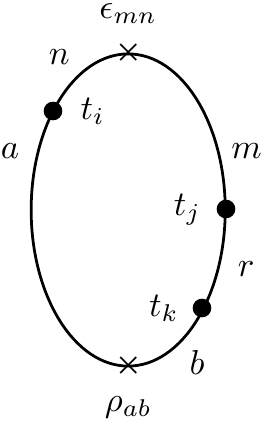}}
\caption{\label{fig:ellipsesx} The 8 graphs corresponding to the third-order matrix element.}
\end{figure}

There is a nice graphical method to compute the matrix element of a general $E_{\cdots}$ sequence. Specifically, for the general state $\rho_{ab}\ket{a}\!\bra{b}$,
we can compute the matrix elements
\begin{equation}
\mathrm{Tr}\Big(\rho_{ab}\ket{a}\!\bra{b} \big[\big[\dots\big[\big[E,M_i\big]_{\eta_i},M_j\big]_{\eta_j},\dots\big],M_N\big]_{\eta_N}\Big)\;
\end{equation}
using the following rules:
\begin{enumerate}
\item Draw $2^N$ clockwise-ordered ellipses with the major axes aligned vertically.

\item Place a cross at the top of the ellipse and associate to it a factor $\epsilon_{mn}$. Place a cross at the bottom of the ellipse and associate to it a factor $\rho_{ab}$.

\item For each of the $N$ times, place a circle on the ellipse. Do this in all possible ways across either the rising (left-hand) or the falling (right-hand) side such that they are always time-ordered vertically, irrespective of their horizontal position.

\item Work clockwise around the ellipse and
\begin{itemize}
\item [(a)] assign a factor of $\mu_{rs}$ for each time,
\item [(b)] connect consecutive times with atom Wightman propagators $\Delta^{r(>)}_{ij}$,
\item [(c)] assign a factor of $e^{+(-)i\omega_{r} t_i}$ for the times $t_i$ followed (preceded) by a cross.
\end{itemize}
\item Assign a factor of $\eta_i$ for any time $t_i$ appearing on the falling side of the ellipse.
\end{enumerate}
These rules are best illustrated by example, and in Figure~\ref{fig:ellipsesx} we show the 8 graphs corresponding to the matrix element of the third-order sequence. Graph (e) in the figure corresponds to the expression
\begin{equation}
	\epsilon_{mn}\, \rho_{ab} \, \mu_{bm} \mu_{ra}\mu_{nr} \, \Delta^{r(>)}_{ij}e^{-i\omega_at_j}e^{i\omega_n t_i}e^{-i\omega_mt_k}e^{i\omega_bt_k}~.
\end{equation} 
When $\epsilon_{mn}=\delta_{mq}\delta_{qn}$ and $\rho_{ab}=\delta_{ag}\delta_{gb}$, the graphs in Figure~\ref{fig:ellipsesx} (a) and (b) are zero, which means that we must place each of the $N$ times such that there is at least one time on each of the rising and falling sides of the ellipse.

\begin{acknowledgements}
This work is supported by the Lancaster-Manchester-Sheffield
Consortium for Fundamental Physics STFC grant ST/J000418/1 and by STFC grant 
ST/L000393/1.
\end{acknowledgements}

\bibliography{atoms}

\begin{thebibliography}{37}%
\makeatletter
\providecommand \@ifxundefined [1]{%
 \@ifx{#1\undefined}
}%
\providecommand \@ifnum [1]{%
 \ifnum #1\expandafter \@firstoftwo
 \else \expandafter \@secondoftwo
 \fi
}%
\providecommand \@ifx [1]{%
 \ifx #1\expandafter \@firstoftwo
 \else \expandafter \@secondoftwo
 \fi
}%
\providecommand \natexlab [1]{#1}%
\providecommand \enquote  [1]{``#1''}%
\providecommand \bibnamefont  [1]{#1}%
\providecommand \bibfnamefont [1]{#1}%
\providecommand \citenamefont [1]{#1}%
\providecommand \href@noop [0]{\@secondoftwo}%
\providecommand \href [0]{\begingroup \@sanitize@url \@href}%
\providecommand \@href[1]{\@@startlink{#1}\@@href}%
\providecommand \@@href[1]{\endgroup#1\@@endlink}%
\providecommand \@sanitize@url [0]{\catcode `\\12\catcode `\$12\catcode
  `\&12\catcode `\#12\catcode `\^12\catcode `\_12\catcode `\%12\relax}%
\providecommand \@@startlink[1]{}%
\providecommand \@@endlink[0]{}%
\providecommand \url  [0]{\begingroup\@sanitize@url \@url }%
\providecommand \@url [1]{\endgroup\@href {#1}{\urlprefix }}%
\providecommand \urlprefix  [0]{URL }%
\providecommand \Eprint [0]{\href }%
\providecommand \doibase [0]{http://dx.doi.org/}%
\providecommand \selectlanguage [0]{\@gobble}%
\providecommand \bibinfo  [0]{\@secondoftwo}%
\providecommand \bibfield  [0]{\@secondoftwo}%
\providecommand \translation [1]{[#1]}%
\providecommand \BibitemOpen [0]{}%
\providecommand \bibitemStop [0]{}%
\providecommand \bibitemNoStop [0]{.\EOS\space}%
\providecommand \EOS [0]{\spacefactor3000\relax}%
\providecommand \BibitemShut  [1]{\csname bibitem#1\endcsname}%
\let\auto@bib@innerbib\@empty
\bibitem [{\citenamefont {Eberhard}\ and\ \citenamefont
  {Ross}(1989)}]{Eberhard:1988yj}%
  \BibitemOpen
  \bibfield  {author} {\bibinfo {author} {\bibfnamefont {P.~H.}\ \bibnamefont
  {Eberhard}}\ and\ \bibinfo {author} {\bibfnamefont {R.~R.}\ \bibnamefont
  {Ross}},\ }\href {\doibase 10.1007/BF00696109} {\bibfield  {journal}
  {\bibinfo  {journal} {Found. Phys.}\ }\textbf {\bibinfo {volume} {2}},\
  \bibinfo {pages} {127} (\bibinfo {year} {1989})}\BibitemShut {NoStop}%
\bibitem [{\citenamefont {Fermi}(1932)}]{Fermi}%
  \BibitemOpen
  \bibfield  {author} {\bibinfo {author} {\bibfnamefont {E.}~\bibnamefont
  {Fermi}},\ }\href@noop {} {\bibfield  {journal} {\bibinfo  {journal} {Rev.
  Mod. Phys.}\ }\textbf {\bibinfo {volume} {4}},\ \bibinfo {pages} {87}
  (\bibinfo {year} {1932})}\BibitemShut {NoStop}%
\bibitem [{\citenamefont {Shirokov}(1967)}]{Shirokov1}%
  \BibitemOpen
  \bibfield  {author} {\bibinfo {author} {\bibfnamefont {M.~I.}\ \bibnamefont
  {Shirokov}},\ }\href@noop {} {\bibfield  {journal} {\bibinfo  {journal} {Sov.
  J. Nucl. Phys.}\ }\textbf {\bibinfo {volume} {4}},\ \bibinfo {pages} {774}
  (\bibinfo {year} {1967})}\BibitemShut {NoStop}%
\bibitem [{Note1()}]{Note1}%
  \BibitemOpen
  \bibinfo {note} {Fermi's original result was consistent with the previous
  work of Kikuchi \cite {Kikuchi} and, despite some initial concerns voiced by
  Ferretti \& Peierls \cite {FP}, was supported by the subsequent work of
  Heitler \& Ma \cite {Heitler} and Hamilton \cite {Hamilton}. In 1968, two
  years after Shirokov \cite {Shirokov1} pointed out Fermi's error, Ferretti
  essentially solved the Fermi problem \cite {Ferretti}. In particular, he
  explained the necessity to focus on a local observable with an inclusive sum
  over unobserved particles. However, Ferretti's work appeared as a chapter in
  a book and seems not to have been widely appreciated (Shirokov was a noteable
  exception \cite {Shirokov2}) and, in the 1970s, Fermi's result was still
  regarded as textbook \cite {Louisell,Milonni}. It is ironic that the Ferretti
  paper starts with the words `In this paper I will not say anything new.' In
  1987, Rubin \cite {Rubin:1987zz} re-discovered the apparently acausal nature
  of the two-atom problem but did not explain how it is resolved. In 1990, the
  correct explanation for the non-violation of Einstein causality in the
  two-atom problem was re-discovered by Biswas et al. \cite {Biswas} and by
  Valentini \cite {Valentini}, and by the mid 1990s the dust seems to have
  settled and the role of the Rotating Wave Approximation in faking causality
  was appreciated \cite {MilonniJamesFearn,PT,Dolce}. A noteable exception to
  this was Hegerfeldt's 1994 paper \cite {Hegerfeldt:1993qe}, which created
  quite a media stir \cite {Maddox,Gribbin} and provoked the clarifying
  response in \cite {Buchholz:1994eb}.}\BibitemShut {Stop}%
  \bibitem [{\citenamefont {Kikuchi}(1930)}]{Kikuchi}%
  \BibitemOpen
  \bibfield  {author} {\bibinfo {author} {\bibfnamefont {S.}~\bibnamefont
  {Kikuchi}},\ }\href@noop {} {\bibfield  {journal} {\bibinfo  {journal}
  {Z.~Phys.}\ }\textbf {\bibinfo {volume} {66}},\ \bibinfo {pages} {558}
  (\bibinfo {year} {1930})}\BibitemShut {NoStop}%
\bibitem [{\citenamefont {Ferretti}\ and\ \citenamefont {Peierls}(1947)}]{FP}%
  \BibitemOpen
  \bibfield  {author} {\bibinfo {author} {\bibfnamefont {B.}~\bibnamefont
  {Ferretti}}\ and\ \bibinfo {author} {\bibfnamefont {R.~E.}\ \bibnamefont
  {Peierls}},\ }\href@noop {} {\bibfield  {journal} {\bibinfo  {journal}
  {Nature (London)}\ }\textbf {\bibinfo {volume} {160}},\
  \bibinfo {pages} {531} (\bibinfo {year} {1947})},\ \bibinfo {note} {{L}etters
  to the Editors}\BibitemShut {NoStop}%
\bibitem [{\citenamefont {Heitler}\ and\ \citenamefont {Ma}(1949)}]{Heitler}%
  \BibitemOpen
  \bibfield  {author} {\bibinfo {author} {\bibfnamefont {W.}~\bibnamefont
  {Heitler}}\ and\ \bibinfo {author} {\bibfnamefont {S.~T.}\ \bibnamefont
  {Ma}},\ }\href@noop {} {\bibfield  {journal} {\bibinfo  {journal} {Proc. R.
  Ir. Acad.}\ }\textbf {\bibinfo {volume} {52}},\ \bibinfo {pages}
  {109} (\bibinfo {year} {1949})}\BibitemShut {NoStop}%
\bibitem [{\citenamefont {Hamilton}(1949)}]{Hamilton}%
  \BibitemOpen
  \bibfield  {author} {\bibinfo {author} {\bibfnamefont {J.}~\bibnamefont
  {Hamilton}},\ }\href@noop {} {\bibfield  {journal} {\bibinfo  {journal}
  {Proc. Phys. Soc. London Sect. A}\ }\textbf {\bibinfo
  {volume} {62}},\ \bibinfo {pages} {12} (\bibinfo {year}
  {1949})}\BibitemShut {NoStop}%
\bibitem [{\citenamefont {Ferretti}(1968)}]{Ferretti}%
  \BibitemOpen
  \bibfield  {author} {\bibinfo {author} {\bibfnamefont {B.}~\bibnamefont
  {Ferretti}},\ }in\ \href@noop {} {\emph {\bibinfo {booktitle} {Old and New
  Problems in Elementary Particles}}},\ \bibinfo {editor} {edited by\ \bibinfo
  {editor} {\bibfnamefont {G.}~\bibnamefont {Puppi}}}\ (\bibinfo  {publisher}
  {Academic Press},\ \bibinfo {address} {New York},\ \bibinfo {year} {1968})\
  p.\ \bibinfo {pages} {108}\BibitemShut {NoStop}%
  \bibitem [{\citenamefont {Shirokov}(1978)}]{Shirokov2}%
  \BibitemOpen
  \bibfield  {author} {\bibinfo {author} {\bibfnamefont {M.~I.}\ \bibnamefont
  {Shirokov}},\ }\href@noop {} {\bibfield  {journal} {\bibinfo  {journal} {Sov.
  Phys. Usp.}\ }\textbf {\bibinfo {volume} {21}},\ \bibinfo {pages} {345}
  (\bibinfo {year} {1978})}\BibitemShut {NoStop}%
\bibitem [{\citenamefont {Louisell}(1973)}]{Louisell}%
  \BibitemOpen
  \bibfield  {author} {\bibinfo {author} {\bibfnamefont {W.~H.}\ \bibnamefont
  {Louisell}},\ }\href@noop {} {\emph {\bibinfo {title} {Quantum Statistical
  Properties of Radiation}}}\ (\bibinfo  {publisher} {Wiley},\ \bibinfo
  {address} {New York},\ \bibinfo {year} {1973})\ p.\ \bibinfo {pages}
  {314}\BibitemShut {NoStop}%
\bibitem [{\citenamefont {Milonni}\ and\ \citenamefont
  {Knight}(1974)}]{Milonni}%
  \BibitemOpen
  \bibfield  {author} {\bibinfo {author} {\bibfnamefont {P.~W.}\ \bibnamefont
  {Milonni}}\ and\ \bibinfo {author} {\bibfnamefont {P.~L.}\ \bibnamefont
  {Knight}},\ }\href@noop {} {\bibfield  {journal} {\bibinfo  {journal} {Phys.
  Rev.}\ }\textbf {\bibinfo {volume} {A10}},\ \bibinfo {pages} {1096} (\bibinfo
  {year} {1974})}\BibitemShut {NoStop}%
\bibitem [{\citenamefont {Rubin}(1987)}]{Rubin:1987zz}%
  \BibitemOpen
  \bibfield  {author} {\bibinfo {author} {\bibfnamefont {M.~H.}\ \bibnamefont
  {Rubin}},\ }\href@noop {} {\bibfield  {journal} {\bibinfo  {journal} {Phys.
  Rev.}\ }\textbf {\bibinfo {volume} {D35}},\ \bibinfo {pages} {3836} (\bibinfo
  {year} {1987})}\BibitemShut {NoStop}%
\bibitem [{\citenamefont {Biswas}\ \emph {et~al.}(1990)\citenamefont {Biswas},
  \citenamefont {Compagno}, \citenamefont {Palma}, \citenamefont {Passante},\
  and\ \citenamefont {Persico}}]{Biswas}%
  \BibitemOpen
  \bibfield  {author} {\bibinfo {author} {\bibfnamefont {A.~K.}\ \bibnamefont
  {Biswas}}, \bibinfo {author} {\bibfnamefont {G.}~\bibnamefont {Compagno}},
  \bibinfo {author} {\bibfnamefont {G.~M.}\ \bibnamefont {Palma}}, \bibinfo
  {author} {\bibfnamefont {R.}~\bibnamefont {Passante}}, \ and\ \bibinfo
  {author} {\bibfnamefont {F.}~\bibnamefont {Persico}},\ }\href {\doibase
  10.1103/PhysRevA.42.4291} {\bibfield  {journal} {\bibinfo  {journal} {Phys.
  Rev.}\ }\textbf {\bibinfo {volume} {A42}},\ \bibinfo {pages} {4291} (\bibinfo
  {year} {1990})},\ \bibinfo {note} {Erratum: Phys. Rev. A{\bf
  44}, 798 (1991)}\BibitemShut {NoStop}%
\bibitem [{\citenamefont {Valentini}(1991)}]{Valentini}%
  \BibitemOpen
  \bibfield  {author} {\bibinfo {author} {\bibfnamefont {A.}~\bibnamefont
  {Valentini}},\ }\href@noop {} {\bibfield  {journal} {\bibinfo  {journal}
  {Phys. Lett.}\ }\textbf {\bibinfo {volume} {A153}},\ \bibinfo {pages} {321}
  (\bibinfo {year} {1991})}\BibitemShut {NoStop}%
\bibitem [{\citenamefont {Power}\ and\ \citenamefont
  {Thirunamachandran}(1997)}]{PT}%
  \BibitemOpen
  \bibfield  {author} {\bibinfo {author} {\bibfnamefont {E.~A.}\ \bibnamefont
  {Power}}\ and\ \bibinfo {author} {\bibfnamefont {T.}~\bibnamefont
  {Thirunamachandran}},\ }\href@noop {} {\bibfield  {journal} {\bibinfo
  {journal} {Phys. Rev.}\ }\textbf {\bibinfo {volume} {A56}},\ \bibinfo {pages}
  {3395} (\bibinfo {year} {1997})}\BibitemShut {NoStop}%
\bibitem [{\citenamefont {Milonni}\ \emph {et~al.}(1995)\citenamefont
  {Milonni}, \citenamefont {James},\ and\ \citenamefont
  {Fearn}}]{MilonniJamesFearn}%
  \BibitemOpen
  \bibfield  {author} {\bibinfo {author} {\bibfnamefont {P.~W.}\ \bibnamefont
  {Milonni}}, \bibinfo {author} {\bibfnamefont {D.~F.~V.}\ \bibnamefont
  {James}}, \ and\ \bibinfo {author} {\bibfnamefont {H.}~\bibnamefont
  {Fearn}},\ }\href@noop {} {\bibfield  {journal} {\bibinfo  {journal} {Phys.
  Rev.}\ }\textbf {\bibinfo {volume} {A52}},\ \bibinfo {pages} {1525} (\bibinfo
  {year} {1995})}\BibitemShut {NoStop}%
\bibitem [{\citenamefont {Dolce}\ \emph {et~al.}(2006)\citenamefont {Dolce},
  \citenamefont {Passante},\ and\ \citenamefont {Persico}}]{Dolce}%
  \BibitemOpen
  \bibfield  {author} {\bibinfo {author} {\bibfnamefont {I.}~\bibnamefont
  {Dolce}}, \bibinfo {author} {\bibfnamefont {R.}~\bibnamefont {Passante}}, \
  and\ \bibinfo {author} {\bibfnamefont {F.}~\bibnamefont {Persico}},\
  }\href@noop {} {\bibfield  {journal} {\bibinfo  {journal} {Phys. Lett.}\
  }\textbf {\bibinfo {volume} {A355}},\ \bibinfo {pages} {152} (\bibinfo {year}
  {2006})}\BibitemShut {NoStop}%
\bibitem [{\citenamefont {Hegerfeldt}(1994)}]{Hegerfeldt:1993qe}%
  \BibitemOpen
  \bibfield  {author} {\bibinfo {author} {\bibfnamefont {G.~C.}\ \bibnamefont
  {Hegerfeldt}},\ }\href {\doibase 10.1103/PhysRevLett.72.596} {\bibfield
  {journal} {\bibinfo  {journal} {Phys. Rev. Lett.}\ }\textbf {\bibinfo
  {volume} {72}},\ \bibinfo {pages} {596} (\bibinfo {year} {1994})}\BibitemShut
  {NoStop}%
\bibitem [{\citenamefont {Maddox}(1994)}]{Maddox}%
  \BibitemOpen
  \bibfield  {author} {\bibinfo {author} {\bibfnamefont {J.}~\bibnamefont
  {Maddox}},\ }\href@noop {} {\bibfield  {journal} {\bibinfo  {journal} {Nature
  (London)}\ }\textbf {\bibinfo {volume}
  {367}},\ \bibinfo {pages} {509} (\bibinfo {year} {1994})},\
  \bibinfo {note} {{N}ews and Views}\BibitemShut {NoStop}%
\bibitem [{\citenamefont {Gribbin}(1994)}]{Gribbin}%
  \BibitemOpen
  \bibfield  {author} {\bibinfo {author} {\bibfnamefont {J.}~\bibnamefont
  {Gribbin}},\ }\href@noop {} {\bibfield  {journal} {\bibinfo  {journal} {New
  Scientist}\ }\textbf {\bibinfo {volume} {1914}},\ \bibinfo {pages} {16}
  (\bibinfo {year} {1994})}\BibitemShut {NoStop}%
  \bibitem [{\citenamefont {Buchholz}\ and\ \citenamefont
  {Yngvason}(1994)}]{Buchholz:1994eb}%
  \BibitemOpen
  \bibfield  {author} {\bibinfo {author} {\bibfnamefont {D.}~\bibnamefont
  {Buchholz}}\ and\ \bibinfo {author} {\bibfnamefont {J.}~\bibnamefont
  {Yngvason}},\ }\href {\doibase 10.1103/PhysRevLett.73.613} {\bibfield
  {journal} {\bibinfo  {journal} {Phys. Rev. Lett.}\ }\textbf {\bibinfo
  {volume} {73}},\ \bibinfo {pages} {613} (\bibinfo {year} {1994})},\ \Eprint
  {http://arxiv.org/abs/hep-th/9403027} {arXiv:hep-th/9403027 [hep-th]}
  \BibitemShut {NoStop}%
\bibitem [{\citenamefont {Schlieder}(1971)}]{Schlieder}%
  \BibitemOpen
  \bibfield  {author} {\bibinfo {author} {\bibfnamefont {S.}~\bibnamefont
  {Schlieder}},\ }in\ \href@noop {} {\emph {\bibinfo {booktitle}
  {Quanten und Felder}}},\ \bibinfo {editor} {edited by\
  \bibinfo {editor} {\bibfnamefont {H.}~\bibnamefont {D{\"u}rr}}}\ (\bibinfo
  {publisher} {Vieweg und Sohn, Verlag, Braunschweig},\
  \bibinfo {year} {1971})\ p.\ \bibinfo {pages} {145}\BibitemShut {NoStop}%
\bibitem [{\citenamefont {Neumann}\ and\ \citenamefont {Werner}(1983)}]{NW}%
  \BibitemOpen
  \bibfield  {author} {\bibinfo {author} {\bibfnamefont {H.}~\bibnamefont
  {Neumann}}\ and\ \bibinfo {author} {\bibfnamefont {R.}~\bibnamefont
  {Werner}},\ }\href@noop {} {\bibfield  {journal} {\bibinfo  {journal} {Int.
  J. Theo. Phys.}\ }\textbf {\bibinfo {volume} {22}},\ \bibinfo {pages} {781}
  (\bibinfo {year} {1983})}\BibitemShut {NoStop}%
\bibitem [{\citenamefont {Hegerfeldt}(1998)}]{Hegerfeldt:1998ar}%
  \BibitemOpen
  \bibfield  {author} {\bibinfo {author} {\bibfnamefont {G.~C.}\ \bibnamefont
  {Hegerfeldt}},\ }\href@noop {} {\bibfield  {journal} {\bibinfo  {journal}
  {Ann. Phys. (Berlin)}\ }\textbf {\bibinfo {volume} {7}},\
  \bibinfo {pages} {716} (\bibinfo {year} {1998})},\ \Eprint
  {http://arxiv.org/abs/quant-ph/9809030} {arXiv:quant-ph/9809030 [quant-ph]}
  \BibitemShut {NoStop}%
\bibitem [{\citenamefont {Cliche}\ and\ \citenamefont
  {Kempf}(2010)}]{Cliche:2009fma}%
  \BibitemOpen
  \bibfield  {author} {\bibinfo {author} {\bibfnamefont {M.}~\bibnamefont
  {Cliche}}\ and\ \bibinfo {author} {\bibfnamefont {A.}~\bibnamefont {Kempf}},\
  }\href {\doibase 10.1103/PhysRevA.81.012330} {\bibfield  {journal} {\bibinfo
  {journal} {Phys. Rev.}\ }\textbf {\bibinfo {volume} {A81}},\ \bibinfo {pages}
  {012330} (\bibinfo {year} {2010})},\ \Eprint {http://arxiv.org/abs/0908.3144}
  {arXiv:0908.3144 [quant-ph]} \BibitemShut {NoStop}%
\bibitem [{\citenamefont
  {Mart{\'i}n-Mart{\'i}nez}(2015)}]{Martin-Martinez:2015psa}%
  \BibitemOpen
  \bibfield  {author} {\bibinfo {author} {\bibfnamefont {E.}~\bibnamefont
  {Mart{\'i}n-Mart{\'i}nez}},\ }\href {\doibase 10.1103/PhysRevD.92.104019}
  {\bibfield  {journal} {\bibinfo  {journal} {Phys. Rev.}\ }\textbf {\bibinfo
  {volume} {D92}},\ \bibinfo {pages} {104019} (\bibinfo {year} {2015})},\
  \Eprint {http://arxiv.org/abs/1509.07864} {arXiv:1509.07864 [quant-ph]}
  \BibitemShut {NoStop}%
\bibitem [{\citenamefont {{Franson}}\ and\ \citenamefont
  {{Donegan}}(2002)}]{Franson}%
  \BibitemOpen
  \bibfield  {author} {\bibinfo {author} {\bibfnamefont {J.~D.}\ \bibnamefont
  {{Franson}}}\ and\ \bibinfo {author} {\bibfnamefont {M.~M.}\ \bibnamefont
  {{Donegan}}},\ }\href {\doibase 10.1103/PhysRevA.65.052107} {\bibfield
  {journal} {\bibinfo  {journal} {\pra}\ }\textbf {\bibinfo {volume} {65}},\
  \bibinfo {eid} {052107} (\bibinfo {year} {2002})},\ \Eprint
  {http://arxiv.org/abs/quant-ph/0108018} {quant-ph/0108018} \BibitemShut
  {NoStop}%
\bibitem [{\citenamefont {Dickinson}\ \emph {et~al.}(2014)\citenamefont
  {Dickinson}, \citenamefont {Forshaw}, \citenamefont {Millington},\ and\
  \citenamefont {Cox}}]{Dickinson:2013lsa}%
  \BibitemOpen
  \bibfield  {author} {\bibinfo {author} {\bibfnamefont {R.}~\bibnamefont
  {Dickinson}}, \bibinfo {author} {\bibfnamefont {J.}~\bibnamefont {Forshaw}},
  \bibinfo {author} {\bibfnamefont {P.}~\bibnamefont {Millington}}, \ and\
  \bibinfo {author} {\bibfnamefont {B.}~\bibnamefont {Cox}},\ }\href {\doibase
  10.1007/JHEP06(2014)049} {\bibfield  {journal} {\bibinfo  {journal} {JHEP}\
  }\textbf {\bibinfo {volume} {06}},\ \bibinfo {pages} {049} (\bibinfo {year}
  {2014})},\ \Eprint {http://arxiv.org/abs/1312.3871} {arXiv:1312.3871
  [hep-th]} \BibitemShut {NoStop}%
\bibitem [{\citenamefont {H{\"u}mmer}\ \emph {et~al.}(2016)\citenamefont
  {H{\"u}mmer}, \citenamefont {Mart{\'i}n-Mart{\'i}nez},\ and\ \citenamefont
  {Kempf}}]{Hummer:2015xaa}%
  \BibitemOpen
  \bibfield  {author} {\bibinfo {author} {\bibfnamefont {D.}~\bibnamefont
  {H{\"u}mmer}}, \bibinfo {author} {\bibfnamefont {E.}~\bibnamefont
  {Mart{\'i}n-Mart{\'i}nez}}, \ and\ \bibinfo {author} {\bibfnamefont
  {A.}~\bibnamefont {Kempf}},\ }\href {\doibase 10.1103/PhysRevD.93.024019}
  {\bibfield  {journal} {\bibinfo  {journal} {Phys. Rev.}\ }\textbf {\bibinfo
  {volume} {D93}},\ \bibinfo {pages} {024019} (\bibinfo {year} {2016})},\
  \Eprint {http://arxiv.org/abs/1506.02046} {arXiv:1506.02046 [quant-ph]}
  \BibitemShut {NoStop}%
\bibitem [{\citenamefont {Schwinger}(1961)}]{Schwinger:1960qe}%
  \BibitemOpen
  \bibfield  {author} {\bibinfo {author} {\bibfnamefont {J.~S.}\ \bibnamefont
  {Schwinger}},\ }\href {\doibase 10.1063/1.1703727} {\bibfield  {journal}
  {\bibinfo  {journal} {J. Math. Phys.}\ }\textbf {\bibinfo {volume} {2}},\
  \bibinfo {pages} {407} (\bibinfo {year} {1961})}\BibitemShut {NoStop}%
\bibitem [{\citenamefont {Keldysh}(1964)}]{Keldysh:1964ud}%
  \BibitemOpen
  \bibfield  {author} {\bibinfo {author} {\bibfnamefont {L.~V.}\ \bibnamefont
  {Keldysh}},\ }\href@noop {} {\bibfield  {journal} {\bibinfo  {journal} {Zh.
  Eksp. Teor. Fiz.}\ }\textbf {\bibinfo {volume} {47}},\ \bibinfo {pages}
  {1515} (\bibinfo {year} {1964})},\ \bibinfo {note} {[Sov. Phys.
  JETP20,1018(1965)]}\BibitemShut {NoStop}%
\bibitem [{\citenamefont {Kobes}\ and\ \citenamefont
  {Semenoff}(1985)}]{Kobes:1985kc}%
  \BibitemOpen
  \bibfield  {author} {\bibinfo {author} {\bibfnamefont {R.~L.}\ \bibnamefont
  {Kobes}}\ and\ \bibinfo {author} {\bibfnamefont {G.~W.}\ \bibnamefont
  {Semenoff}},\ }\href {\doibase 10.1016/0550-3213(85)90056-2} {\bibfield
  {journal} {\bibinfo  {journal} {Nucl. Phys.}\ }\textbf {\bibinfo {volume}
  {B260}},\ \bibinfo {pages} {714} (\bibinfo {year} {1985})}\BibitemShut
  {NoStop}%
\bibitem [{\citenamefont {Kobes}\ and\ \citenamefont
  {Semenoff}(1986)}]{Kobes:1986za}%
  \BibitemOpen
  \bibfield  {author} {\bibinfo {author} {\bibfnamefont {R.~L.}\ \bibnamefont
  {Kobes}}\ and\ \bibinfo {author} {\bibfnamefont {G.~W.}\ \bibnamefont
  {Semenoff}},\ }\href {\doibase 10.1016/0550-3213(86)90006-4} {\bibfield
  {journal} {\bibinfo  {journal} {Nucl. Phys.}\ }\textbf {\bibinfo {volume}
  {B272}},\ \bibinfo {pages} {329} (\bibinfo {year} {1986})}\BibitemShut
  {NoStop}%
\bibitem [{\citenamefont {Cutkosky, R.
  E.}(1960)}]{Cutkosky:1960sp}%
  \BibitemOpen
  \bibfield  {author} {\bibinfo {author} {\bibnamefont
  {Cutkosky, R. E.}},\ }\href {\doibase 10.1063/1.1703676}
  {\bibfield  {journal} {\bibinfo  {journal} {J. Math. Phys.}\ }\textbf
  {\bibinfo {volume} {1}},\ \bibinfo {pages} {429} (\bibinfo {year}
  {1960})}\BibitemShut {NoStop}%
\bibitem [{\citenamefont {'t~Hooft}\ and\ \citenamefont
  {Veltman}(1974)}]{'tHooft:1973pz}%
  \BibitemOpen
  \bibfield  {author} {\bibinfo {author} {\bibfnamefont {G.}~\bibnamefont
  {'t~Hooft}}\ and\ \bibinfo {author} {\bibfnamefont {M.~J.~G.}\ \bibnamefont
  {Veltman}},\ }\bibfield  {booktitle} {\emph {\bibinfo {booktitle} {{2nd
  Summer Institute on Particle Interactions at Very High Energies: Duality in
  Elementary Particle Physics Louvain-la-Neuve, Belgium}}},\ }\href@noop {}
  {\bibfield  {journal} {\bibinfo  {journal} {NATO Sci. Ser. B}\ }\textbf
  {\bibinfo {volume} {4}},\ \bibinfo {pages} {177} (\bibinfo {year}
  {1974})}\BibitemShut {NoStop}%
\bibitem [{\citenamefont {Kobes}(1991)}]{Kobes:1990ua}%
  \BibitemOpen
  \bibfield  {author} {\bibinfo {author} {\bibfnamefont {R.}~\bibnamefont
  {Kobes}},\ }\href {\doibase 10.1103/PhysRevD.43.1269} {\bibfield  {journal}
  {\bibinfo  {journal} {Phys. Rev.}\ }\textbf {\bibinfo {volume} {D43}},\
  \bibinfo {pages} {1269} (\bibinfo {year} {1991})}\BibitemShut {NoStop}%
\end{thebibliography}%

\end{document}